\documentstyle{mn}
\input{psfig}
\newif\ifAMStwofonts

\def\simgt{\hbox{\rlap{\raise 0.425ex\hbox{$>$}}\lower 0.65ex\hbox{$\sim$}}}
\def\simlt{\hbox{\rlap{\raise 0.425ex\hbox{$<$}}\lower 0.65ex\hbox{$\sim$}}}

\def\kms{km~s$^{-1}$}
\def\bj{b_{\rm J}}
\def\CIV{C{\small~IV}}
\def\CIII{C{\small~III]}}
\def\MgII{Mg{\small~II}}
\def\FeII{Fe{\small~II}}
\def\CaII{Ca{\small~II}}
\def\AlII{Al{\small~II}}
\def\AlIII{Al{\small~III}}
\def\OII{[O{\small~II}]}

\ifoldfss
  \ifCUPmtlplainloaded \else
    \NewTextAlphabet{textbfit} {cmbxti10} {}
    \NewTextAlphabet{textbfss} {cmssbx10} {}
    \NewMathAlphabet{mathbfit} {cmbxti10} {} 
    \NewMathAlphabet{mathbfss} {cmssbx10} {} 
  \fi
  \ifAMStwofonts
    \ifCUPmtlplainloaded \else
      \NewSymbolFont{upmath} {eurm10}
      \NewSymbolFont{AMSa} {msam10}
      \NewMathSymbol{\upi}     {0}{upmath}{19}
      \NewMathSymbol{\umu}     {0}{upmath}{16}
      \NewMathSymbol{\upartial}{0}{upmath}{40}
      \NewMathSymbol{\leqslant}{3}{AMSa}{36}
      \NewMathSymbol{\geqslant}{3}{AMSa}{3E}

      \let\leq=\leqslant 
      \let\geq=\geqslant \let\ge=\geqslant
    \fi
  \fi
\fi 

\ifnfssone
  \newmathalphabet{\mathit}
  \addtoversion{normal}{\mathit}{cmr}{m}{it}
  \addtoversion{bold}{\mathit}{cmr}{bx}{it}
  \newmathalphabet{\mathbfit} 
  \addtoversion{normal}{\mathbfit}{cmr}{bx}{it}
  \addtoversion{bold}{\mathbfit}{cmr}{bx}{it}
  \newmathalphabet{\mathbfss} 
  \addtoversion{normal}{\mathbfss}{cmss}{bx}{n}
  \addtoversion{bold}{\mathbfss}{cmss}{bx}{n}
  \ifAMStwofonts
    \ifCUPmtlplainloaded \else
      \UseAMStwoboldmath
      \makeatletter
      \new@mathgroup\upmath@group
      \define@mathgroup\mv@normal\upmath@group{eur}{m}{n}
      \define@mathgroup\mv@bold\upmath@group{eur}{b}{n}
      \edef\UPM{\hexnumber\upmath@group}
      \new@mathgroup\amsa@group
      \define@mathgroup\mv@normal\amsa@group{msa}{m}{n}
      \define@mathgroup\mv@bold\amsa@group{msa}{m}{n}
      \edef\AMSa{\hexnumber\amsa@group}
      \makeatother
      \mathchardef\upi="0\UPM19
      \mathchardef\umu="0\UPM16
      \mathchardef\upartial="0\UPM40
      \mathchardef\leqslant="3\AMSa36
      \mathchardef\geqslant="3\AMSa3E

      \let\leq=\leqslant 
      \let\geq=\geqslant \let\ge=\geqslant
    \fi
  \fi
\fi 

\ifnfsstwo
  \DeclareMathAlphabet{\mathbfit}{OT1}{cmr}{bx}{it}
  \SetMathAlphabet\mathbfit{bold}{OT1}{cmr}{bx}{it}
  \DeclareMathAlphabet{\mathbfss}{OT1}{cmss}{bx}{n}
  \SetMathAlphabet\mathbfss{bold}{OT1}{cmss}{bx}{n}
  \ifAMStwofonts
    \ifCUPmtlplainloaded \else
      \DeclareSymbolFont{UPM}{U}{eur}{m}{n}
      \SetSymbolFont{UPM}{bold}{U}{eur}{b}{n}
      \DeclareSymbolFont{AMSa}{U}{msa}{m}{n}
      \DeclareMathSymbol{\upi}{0}{UPM}{"19}
      \DeclareMathSymbol{\umu}{0}{UPM}{"16}
      \DeclareMathSymbol{\upartial}{0}{UPM}{"40}
      \DeclareMathSymbol{\leqslant}{3}{AMSa}{"36}
      \DeclareMathSymbol{\geqslant}{3}{AMSa}{"3E}

      \let\leq=\leqslant 
      \let\geq=\geqslant \let\ge=\geqslant
    \fi
  \fi
\fi 

\ifCUPmtlplainloaded \else
  \ifAMStwofonts \else 
    \def\upi{\pi}
    \def\umu{\mu}
    \def\upartial{\partial}
  \fi
\fi
\title[The 2dF BL Lac Survey]{The 2dF BL Lac Survey}
\author[D.Londish et al.]
	{D. Londish$^{1,2}$, S.M. Croom$^{2}$, B.J. Boyle$^{2}$,
   T. Shanks$^3$, P.J. Outram$^3$, E.M. Sadler$^{1}$,\newauthor
  N.S. Loaring$^4$, R.J. Smith$^5$, L. Miller$^4$, P.F.L. Maxted$^{6,7}$\\
${^1}$ University of Sydney, School of Physics, Sydney NSW 2006, Australia \\
${^2}$ Anglo-Australian Observatory, PO Box 296, Epping, NSW 1710, Australia\\
${^3}$ Department of Physics, University of Durham, South Road, 
Durham, DH1 3LE, UK\\
${^4}$ Department of Physics, Oxford University, 1 Keble Road, Oxford,
OX1 3RH, UK \\
${^5}$ Liverpool John Moores University, Twelve Quays House, Egerton
Wharf, Birkenhead, CH41 1LD, UK\\
${^6}$ Department of Physics \& Astronomy, University of Southampton,
  Highfield, Southampton SO17 1BJ, UK\\
${^7}$Department of Physics, Keele University, Staffordshire, ST5 5BG, UK} 

\begin{document}

\maketitle

\newcommand{\fmmm}[1]{\mbox{$#1$}}
\newcommand{\scnd}{\mbox{\fmmm{''}\hskip-0.3em .}}
\newcommand{\scnp}{\mbox{\fmmm{''}}}

\begin{abstract}
We have optically identified a sample of 56 featureless continuum
objects without significant proper motion from the 2dF QSO Redshift
Survey (2QZ).  The steep number--magnitude relation of the sample,
$n(\bj) \propto 10^{0.7\bj}$, is similar to that derived for QSOs in
the 2QZ and inconsistent with any population of Galactic objects.
Follow up high resolution, high signal-to-noise, spectroscopy of five
randomly selected objects confirms the featureless nature of these
sources.  Assuming the objects in the sample to be largely featureless
AGN, and using the QSO evolution model derived for the 2QZ, we predict
the median redshift of the sample to be $z=1.1$. This model also
reproduces the observed number-magnitude relation of the sample using
a renormalisation of the QSO luminosity function, $\Phi^* =
\Phi^*_{\rm \sc qso}/66 \simeq 1.65 \times
10^{-8}\,$mag$^{-1}$Mpc$^{-3}$.  Only $\sim$20 per cent of the objects
have a radio flux density of $S_{1.4}>3\,$mJy, and further VLA
observations at 8.4 GHz place a  $5\sigma$ limit of
$S_{8.4} < 0.2$mJy 
on the bulk of the sample. We postulate that these objects could form
a population of radio-weak AGN with weak or absent emission lines,
whose optical spectra are indistinguishable from those of BL Lac
objects.
\end{abstract}

\section{Introduction}

Over the last decade research into the nature of BL Lac objects has
greatly increased our understanding of this enigmatic and
intrinsically rare class of active galactic
nuclei (AGN). Nevertheless, small number statistics and inhomogeneous
samples have left many key questions unanswered.  In particular the
evolutionary behaviour of the BL Lac population is poorly known, and
from recent survey results (e.g. Padovani \& Giommi 1995; Bade et al. 
1998; Rector et al. 2000) appears inconsistent with trends observed in
the population of AGN as a whole.

BL Lacs are thought to be dominated by emission from a non-thermal,
relativistic jet emanating from the galaxy's core.  Their optical
spectra are thus characterised by an absence of prominent emission or
absorption features, making them difficult to target in optical
surveys. Initial detection is therefore usually at either X-ray or
radio frequencies, with subsequent optical follow-up. Not
surprisingly, surveys conducted at these different frequencies find
differing classes of objects, with the BL Lacs selected in radio
surveys (e.g. the 1 Jy BL Lac Sample, Stickel et al. 1991) being
predominantly the radio-loud or low energy peaked variety (LBLs),
whilst X-ray selected samples are characterised by a population of
less luminous (optically and at radio frequencies), X-ray dominant,
high energy peaked BL Lacs (HBLs) (e.g. the EMSS BL Lac Sample, Morris
et al. 1991).

Although deeper, multiwavelength surveys have overcome some of the
problems of selection bias and high flux limits (for example, the
$ROSAT$ All Sky Survey-Greenbank BL Lac Sample (Laurent-Muehleisen et
al. 1999), revealing a new population of IBLs -- BL Lacs with
properties intermediate between those of LBLs and HBLs), greater
coverage of objects at the very faint end of the luminosity scale is
still necessary if the question of BL Lac evolution is to be addressed
satisfactorily. The current understanding is that, while LBLs and IBLs
exhibit mildly positive or no evolution (Rector \& Stocke 2001,
Laurent-Muehleisen et al. 1999) the HBLs appear to evolve negatively,
that is they were either less numerous or less luminous in the past
(Wolter et al. 1994, Rector et al. 2000). This is in direct contrast to
the trend of strong, positive evolution exhibited by other classes of
AGN.

Estimating the relative numbers of LBLs and HBLs is similarly hampered
by survey flux limits and selection biases (Urry \& Padovani 1995;
Fossati 2001).  Furthermore finding low luminosity BL Lac objects
at higher redshifts is likely to be difficult at X-ray frequencies,
given that these objects are also characterised by a lower Doppler
factor, $\delta$\footnote{$\delta = (\Gamma(1 - \beta
cos\theta))^{-1}$, where $\Gamma = (1 - \beta^2)^{-1/2}$, $\theta$ is
the angle to the line of sight and $\beta c$ is the bulk velocity of
the jet; when $\theta = 1/ \Gamma, \delta = \Gamma$}, and hence lower
$\nu_{peak}$ (Giommi et al. 2001). Thus at the emitted frequency the
flux may well fall below the survey flux limit, and as a consequence
omission of these fainter, extreme HBLs would result in an artificial
flattening of the number--flux (N--S) relation for this class of BL Lac. 

Clearly selection at optical wavelengths (below $\nu_{peak}$) would
overcome some of the bias in estimating numbers of LBLs and HBLs (Urry
\& Padovani 1995). Furthermore, an optically flux-limited survey
essentially places no upper or lower limits on the radio and X-ray
fluxes of objects in the sample; the difficulty has always been the
amount of telescope time required to assemble such a sample. Recently,
however, an opportunity has arisen to select BL Lac candidates
primarily from optical observations using the 2dF QSO Redshift Survey
(2QZ, Croom et al. 2001).  On completion the 2QZ will contain almost
50000 spectra of $UBR$ colour-selected stellar objects at magnitudes
$18.25< \bj \leq 20.85$, of which approximately 25000 will
be QSOs.

In this paper we present an initial analysis of 56 objects with
featureless continuum spectra and no significant proper motion,
selected from the 2QZ catalogue as at July 2001.  Our procedure for
the selection of BL Lac candidates from this catalogue is described in
section 2; in section 3 we present details of the radio and optical 
observations and derive in section 4 the number-magnitude relation for
this sample of candidate BL Lac objects.  A discussion of our
findings follows in Section 5.

\section{The 2dF BL Lac Survey}

\subsection{The 2QZ catalogue}

The 2QZ is a large spectroscopic survey of QSO candidates which, when
complete, will result in the spectroscopic identification of a
catalogue of $\sim $25000 QSOs, more than twenty times larger than any
previous such catalogue. The survey has been made possible by the
2-degree field (2dF), 400-fibre multi-object spectrograph at the
Anglo-Australian Telescope (Lewis et al. 2001).  The survey area covers
a total of 740 deg$^2$, made up of two 75$^o$ x 5$^o$ strips, one
centered on $\delta = \ -30^{\circ}$ with RA range 21$^h$40 to 3$^h$15
(southern strip) and the other centered on $\delta = \ 0^o$ with RA
range 9$^h$50 to 14$^h$50 (equatorial strip).  QSO candidates were
selected from APM measurements of UK Schmidt $U, \ J$ and $R$
plates/films on the basis of their $u-\bj/\bj-r$ colours and 18.25 $<
\bj \leq 20.85$.  A detailed description of the creation of the input
catalogue is given in Smith et al. (2002). The key features
of the selection criteria are that objects satisfy either $u'-\bj \leq
-0.36$ or $u'-\bj < 0.12-0.8(\bj-r)$ or $\bj-r < 0.05$, where
$u'=u-0.24$.  Further information is available at http://www.2dfquasar.org. 

\subsection{Selection of continuum sources}

The first 2dF observations for the QSO survey were made in September
1997 and by July 2001 the catalogue contained 40736 objects, of which
50.5 per cent are QSOs. This version of the catalogue, representing an
intermediate stage between the 10k catalogue (Croom et al.\ 2001) and
final catalogues, was used as the basis for selecting the BL Lac
candidates in the current sample. The total area covered by this
catalogue is 613 deg$^2$.

The spectroscopic data from 2dF were initially reduced using the 2dF
pipeline reduction system (Bailey et al. 2002).  Classification and
redshift estimation for the QSOs was carried out using a specially
written ${\sc autoz}$ code (Miller et al.\ in preparation) based on a
$\chi^2$ fit to stellar/QSO templates.  All ${\sc autoz}$
classifications were checked and double-checked by independent 2QZ
team members to ensure a high-level of catalogue reliability (97 per
cent, see Croom et al.\ 2001).  With the express purpose of
identifying objects with featureless spectra, all spectra satisfying
$18.25<\bj\leq 20.00$ and with a measured signal-to-noise ratio (SNR)
of $\geq10.0$ were  visually re-examined (see section 2.3).  Fainter
objects satisfying the SNR criterion were not included in the analysis
because it would be difficult to obtain follow-up spectroscopic
observations at higher resolution/signal-to-noise.  A number of fields
were observed during periods of significant moonlight.  In some cases
this moonlight caused excess scattered light in the 2dF spectrographs
which was not completely removed by the data reduction process. This
produced artificially high, featureless continuum levels for some
objects observed in those fields.  Consequently we removed all fields
with strong moonlight, determined from sky counts in that field.  With
the removal of these fields the total survey area became 544 deg$^2$.

\subsection{Visual classification}

From this area we selected for visual re-examination 6772 objects
satisfying the above selection criteria.  Approximately 15 per cent of
the objects in the 2QZ have two or more spectra obtained at different
times during survey observations.  Different spectra of the same
objects were also inspected and classified independently.

Objects with featureless continua are inherently difficult to
classify.  For consistency, we adopted the standard operational
definition of a BL Lac as an extragalactic object with no emission
lines greater than $W_{\lambda}=$5\AA\ and \CaII\ H\&K break
contrast of less than 0.25 (Stocke et al.\ 1990). We took great care
to remove all Galactic subdwarfs with weak absorption lines (\CaII\
H\&K, Balmer series, He{\small~I},  He{\small~II}, Mg{\small~H})
and also 
QSOs with emission lines.

Provided they met these criteria, we also included objects with
non-stellar absorption lines to avoid bias against high redshift
objects with intervening absorbers.  We also included objects with
extremely broad ($>200$\AA) and unidentifiable emission humps or dips
in the continuum.  This was to avoid bias against lineless QSOs with
broad absorption line systems which would also fulfil the BL Lac
classification criteria.  Care was taken to avoid selecting peculiar
low-ionisation broad absorption line (BAL) QSOs (Becker et al.\ 1997)
under this latter criterion.  We attempted to minimise any
subjectivity in this visual classification scheme by using three
independent examiners (DL, SMC and BJB).  To be accepted as a continuum
source, an object spectrum has to be classified as such by at least
two people independently.  Two of the people who carried out the
visual inspection (SMC and BJB) have also examined visually all 40000
spectra observed to date (August 2001) in the 2QZ.  Selection of continuum
objects was therefore made by those with significant experience of the 
visual appearance of a wide range of optical QSO spectra. While the
'continuum' spectra selected here may simply represent the objects in
the extreme tail of the QSO emission line equivalent width distribution,
they are markedly different in spectral appearance from the 20000 QSOs
identified to date in the 2QZ.

Based on these criteria, 74 objects out of the 6772
initially selected to have high signal-to-noise spectra were
classified as having continuum spectra.

\subsection{Removing contamination by white dwarfs}

Although the identifying feature of a BL Lac object at optical
wavelengths is an absence of features, it clearly does not follow that
all featureless continuum objects are BL Lacs. The most likely non-AGN
class of objects which could be confused with a BL Lac on the basis of
their optical spectra are continuum DC white dwarfs (e.g. Wesemael et
al 1993).  One feature which can distinguish between AGN and DC white
dwarfs is a measurement of proper motion.  We would expect DC white
dwarfs to exhibit a detectable proper motion over the typical period
between the UKST $J$ and $R$ photographic plates used to create the
2QZ input catalogue.  Based on the McCook \& Sion (1999) catalogue DC
white dwarfs exhibit a range in absolute magnitude $12 < M_B < 16$
with a median $M_B = 14$. Even if we assume that the colour selection
would have biased any selection of  DC white dwarfs towards the
brightest absolute magnitudes ($M_B\sim 12$), this corresponds to a
distance of 275 pc at $B=19.2$, the median  magnitude of our
sample. With typical transverse velocities observed in the  range
$30-80$ \kms ~(Sion et al. 1988), at this distance DC white dwarfs
would exhibit proper motions $\sim23-60$ mas yr$^{-1}$.

We used the SuperCOSMOS Sky Survey\footnote {SuperCosmos Sky Survey at
www-wfau.roe.ac.uk/sss, maintained by the Institute for Astronomy,
Royal Observatory, Edinburgh.} to search for proper motions in our
sample based on the positions measured from the UKST $J$ and $R$ sky
survey plates.  Given the typical astrometric accuracy obtainable from
the $J$ and $R$ photographic plates ($\sim0.10$ arcsec in each co-ordinate)
and the typical mean epoch separation (16 years for fields in the southern strip,
and 11 years for 8 fields in the equatorial strip), we are sensitive
at the $3\sigma$ level to proper motions with 30 mas\,yr$^{-1}$.  This
should be sufficient to detect the bulk of the DC white dwarf
population.
\begin{figure}
{\hspace*{0.3in}{\psfig{file=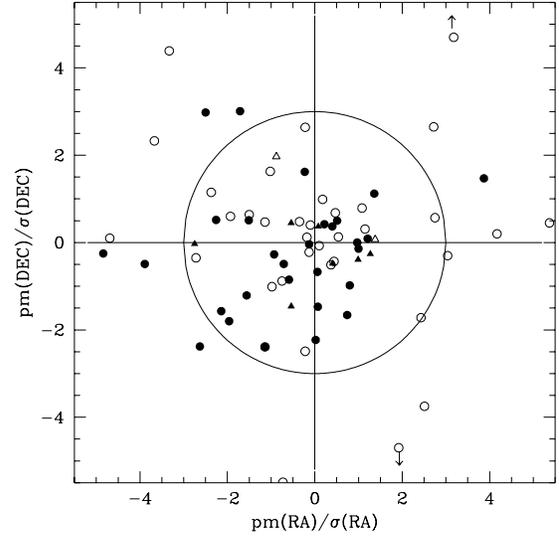,height=3in}}}
\caption{Proper motions for the 74 featureless continuum objects in
the sample; filled triangles/circles represent objects with/without
detectable radio flux and for which $\sigma <15\,$mas$\,$yr$^{-1}$ in
each co-ordinate, while open circles/open triangles are sources with $
\sigma >$15  mas/yr. The circle represents the 3$\sigma$
cutoff. Objects outside the circle have been removed from the sample.}
\label{pm}
\end{figure}

\begin{figure}
{\hspace*{0.3in}{\psfig{file=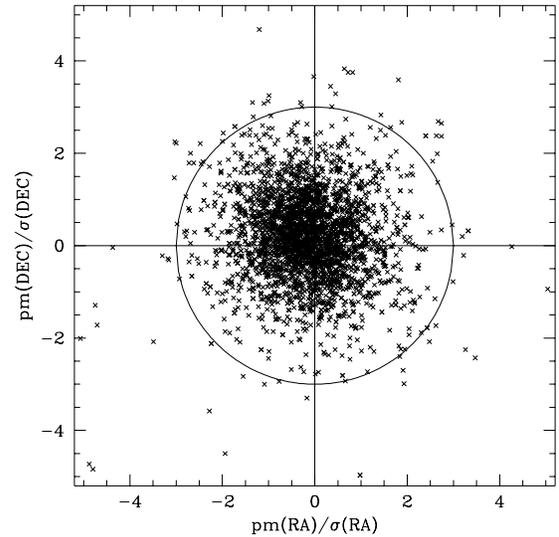,height=3in}}}
\caption{Proper motions for all 2QZ QSOs with 18.25 $< \bj < 19.75$
and SNR $\geq10$. About 2 per cent of sources lie outside the
3$\sigma$ circle and can be attributed to plate inaccuracies. }
\label{qpm}
\end{figure}

\begin{table*}
\caption{Objects in the 2BL sample.}
\label{table_bl}
\begin{center}
\begin{tabular}{|l|l|r|c|r|r|r|r|r|r|l|} \hline
\multicolumn{1}{|c|} {$ \ $ } &
\multicolumn{1}{|c|} {$ \ $ } &
\multicolumn{1}{|c|} {$ \ $ } &
\multicolumn{1}{|c|} {$ \ $ } &
\multicolumn{1}{|c|} {$ \ $ } &
\multicolumn{1}{|c|} {$ \ $ } &
\multicolumn{1}{|c|} {$S_{1.4 }$ } &
\multicolumn{1}{|c|} {$S_{8.4 }$ } &
\multicolumn{1}{|c|} {$f_x \times 10^{-12}$} &
\multicolumn{1}{|c|} {$ \ $ } &
\multicolumn{1}{|c|} {$ \ $ } \\
\multicolumn{1}{|c|} {object} &
\multicolumn{1}{|c|}  {RA (2000)} & 
\multicolumn{1}{|c|} {dec (2000)} &
\multicolumn{1}{|c|} {$\bj$} &
\multicolumn{1}{|c|} {$u-\bj$ } &
\multicolumn{1}{|c|} {$\bj-r$ } &
\multicolumn{1}{|c|} {mJy} &
\multicolumn{1}{|c|} {mJy} &
\multicolumn{1}{|c|} {\footnotesize erg$s^{-1}$cm$^{-2}$ } &
\multicolumn{1}{|c|} {$z$ } &
\multicolumn{1}{|c|} {Notes } \\
\hline
J002522.8$-$284034 & 00  25  22.80 & $-$28  40  34.9 & 18.95 & $-$1.33 & $-$0.12 &   $<$2.8   &        &      &         &  \\  
J002746.6$-$293308 & 00  27  46.66 & $-$29  33  08.4 & 19.08 & $-$1.09 & $-$0.03 &  $<$3.0    &        &      &         &  \\  
J002751.5$-$293506 & 00  27  51.56 & $-$29  35  06.9 & 18.83 & $-$0.84
&   0.16 & $<$2.8    &  $<$0.2 &      &         & f  \\  
J003058.2$-$275629 & 00  30  58.27 & $-$27  56  29.8 & 19.03 & $-$0.73 &   0.27 &    $<$2.8   &  $<$0.2 &      & 1.577   &  \\  
J004106.8$-$291114 & 00  41  06.83 & $-$29  11  14.7 & 18.97 & $-$0.71 & $<-$2.08 &    $<$2.8   &  $<$0.2 &      &         &\\  
J004950.6$-$284907 & 00  49  50.69 & $-$28  49  07.9 & 19.70 & $-$0.53 &   0.52 &    $<$2.8  &        &      &         &  \\  
J014310.1$-$320056 & 01  43  10.10 & $-$32  00  56.7 & 19.36 & $-$0.90 &   1.65 & 76.0 &        & 0.42 &         &a \\  
J023405.5$-$301519 & 02  34  05.58 & $-$30  15  19.5 & 18.54 & $-$0.95 &   0.57 &    $<$2.8   &        &      & 1.710   &  \\  
J023536.7$-$293843 & 02  35  36.70 & $-$29  38  43.6 & 18.64 & $-$0.25 & $-$0.63 & 5.0  &        & 1.0  &         &b \\  
J024659.5$-$294822 & 02  46  59.58 & $-$29  48  22.6 & 19.90 & $-$0.08 &   0.04 &   $<$3.0   &        &      &         &  \\  
J030416.3$-$283217 & 03  04  16.33 & $-$28  32  17.9 & 19.54 & $-$0.69 &   0.52 &
8.0  &        & 2.2  &         &c,f \\  
J031056.9$-$305901 & 03  10  56.91 & $-$30  59  01.3 & 18.70 & $-$0.67 &   0.37 &  $<$3.0    &        &      & $>$0.70 &  \\  
J100253.2$-$001728 & 10  02  53.27 & $-$00  17  28.6 & 19.20 & $-$0.91 &   0.17 &    $<$1.0   &  $<$0.2 &      &         &e \\  
J102615.3$-$000630 & 10  26  15.34 & $-$00  06  30.6 & 19.40 & $-$0.98
&   0.23 &    $<$1.0 &        &      &         &e \\  
J103607.4+015658  & 10  36  07.48 &  +01  56  58.4 & 19.23 & $-$0.94 &
0.66 &   $<$1.0 &        &      &         &  \\  
J104519.7+002615  & 10  45  19.71 &  +00  26  15.0 & 18.68 & $-$1.03 &   0.25 &
$<$1.0 &        &      &         & e,f \\
J105355.1$-$005538 & 10  53  55.18 & $-$00  55  38.5 & 19.53 & $-$0.77 &   0.44 & $<$1.0  &        &      &         &  f\\  
J105534.3$-$012617 & 10  55  34.36 & $-$01  26  17.3 & 18.96 & $-$1.22 &   0.46 & 11.0 &        & 0.33 &         &d,e,f\\  
J110644.5+000717  & 11  06  44.52 &  +00  07  17.4 & 19.86 & $-$1.11 & $< -$1.08 & $<$1.0  &        &      &         &\\  
J113413.4+001041  & 11  34  13.47 &  +00  10  41.3 & 18.80 & $-$0.74 &   0.55 & $<$1.0  &        &      &         &  \\  
J113900.5$-$020140 & 11  39  00.54 & $-$02  01  40.9 & 19.68 & $-$0.48 &   0.92 & $<$1.0 &        &      & $>$0.61 &  \\  
J114010.5$-$002936 & 11  40  10.51 & $-$00  29  36.4 & 19.90 & $-$1.02 & $-$0.07 & $<$1.0 &        &      &         &  \\  
J114137.1$-$002730 & 11  41  37.10 & $-$00  27  30.8 & 19.97 & $-$1.55 &   1.26 &  $<$1.0 &        &      &         &  \\  
J114221.4$-$014812 & 11  42  21.42 & $-$01  48  12.2 & 19.30 & $-$1.28 &   0.89 &  $<$1.0 &        &      & 1.276   &  \\  
J114327.3$-$005050 & 11  43  27.30 & $-$00  50  50.6 & 19.97 & $-$0.39 &   0.36 & $<$1.0  &        &      & 1.591   &  \\  
J114521.6$-$024758 & 11  45  21.61 & $-$02  47  58.3 & 18.79 & $-$0.95 &   0.26 &   $<$1.0    &  $<$0.2 &      &         &  \\  
J114554.8+001023  & 11  45  54.85 &  +00  10  23.6 & 19.59 & $-$0.83 &   0.13 & $<$1.0  &        &      &         &  \\  
J115909.6$-$024534 & 11  59  09.61 & $-$02  45  34.9 & 19.24 & $-$0.81 &   0.47 & $<$1.0  &        &      &         &  \\  
J120015.3+000552  & 12  00  15.35 &  +00  05  52.6 & 19.81 & $-$0.51 &   1.18 &  $<$1.0 &        &      & $>$0.94 &  \\  
J120558.1$-$004216 & 12  05  58.17 & $-$00  42  16.3 & 19.11 & $-$0.47 &   0.65 & $<$1.0  &        &      &         & f \\  
J120801.8$-$004219 & 12  08  01.85 & $-$00  42  19.5 & 18.96 & $-$1.01
& 0.10 &     $<$1.0  & $<0.2 $ &       &         &  f \\  
J121834.8$-$011955 & 12  18  34.88 & $-$001  19  55.9 & 19.70 & $-$1.02 &   1.45 & 244.0&        &      &         &e \\  
J122338.0$-$015619 & 12  23  38.05 & $-$01  56  19.1 & 19.22 & $-$1.16 &   0.37 &   $<$1.0        &      &         &  \\  
J123437.6$-$012953 & 12  34  37.64 & $-$01  29  53.1 & 19.44 & $-$0.64 &   0.47 &  $<$1.0 &        &      & $>$1.06 &  \\  
J125435.7$-$011822 & 12  54  35.76 & $-$01  18  22.5 & 19.44 & $-$0.63 & 0.63 & $<$1.0 &        &      &         &  \\  
J130009.8$-$022601 & 13  00  09.89 & $-$02  26  01.4 & 19.22 & $-$0.66 &   0.55 &  $<$1.0 &        &      &         &  \\  
J131635.1$-$002810 & 13  16  35.14 & $-$00  28  10.6 & 19.82 & $-$0.36 &   0.29 &  $<$1.0&        &      &         &  f\\  
J132811.5+000227  & 13  28  11.54 &  +00  02  27.8 & 19.79 & $-$0.59 &   0.00 &  $<$1.0&        &      &         &  \\  
J140021.0+001955  & 14  00  21.06 &  +00  19  55.9 & 19.89 & $-$0.97 &   0.59 &  $<$1.0 &        &      &         &  \\  
J140207.7$-$013033 & 14  02  07.70 & $-$01  30  33.3 & 19.71 & $-$0.71 & $-$0.43 &  $<$1.0&        &      &         &  f \\  
J140916.3$-$000012 & 14  09  16.36 & $-$00  00  12.0 & 18.84 & $-$0.60 &   0.37 &  $<$1.0 &        &      &         & f  \\  
J141040.2$-$023020 & 14  10  40.28 & $-$02  30  20.7 & 19.43 & $-$1.03 &   0.16 &    $<$1.0   &  $<$0.2 &      &         & f  \\  
J142526.2$-$011826 & 14  25  26.20 & $-$01  18  26.3 & 19.91 & $-$0.26 &   0.44 & 10.0 &        &      & 0.041   &  \\  
J215454.3$-$305654 & 21  54  54.35 & $-$30  56  54.3 & 19.55 & $-$0.78 &   0.66 &   $<$2.8    &  $<$0.2 &      &         &  \\  
J220515.8$-$311537 & 22  05  15.84 & $-$31  15  37.5 & 19.58 & $-$0.60 &   0.30 &  $<$2.5 &        &      &         &  f \\  
J220850.0$-$302817 & 22  08  50.02 & $-$30  28  17.6 & 18.40 & $-$0.77 &   0.37 &  $<$2.8 &        &      &         &  \\  
J221105.2$-$284933 & 22  11  05.25 & $-$28  49  33.0 & 18.57 & $-$0.70 &   0.48 & 81.0 &        & 0.26 & $>$1.85 & g \\  
J221450.1$-$293225 & 22  14  50.11 & $-$29  32  25.2 & 19.21 & $-$0.37
&   0.79 &   $<$3.0 &        &      &         &  \\  
J223233.5$-$272859 & 22  32  33.57 & $-$27  28  59.9 & 19.33 & $-$1.00 & $-$0.10 &  $<$3.0 &        &      &         &  \\  
J224559.1$-$312223 & 22  45  59.10 & $-$31  22  23.3 & 19.38 & $-$0.49 &   0.66 &  $<$3.0 &        &      &         &  f \\  
J225453.2$-$272509 & 22  54  53.20 & $-$27  25  09.4 & 18.83 & $-$1.31 &   2.01 & 52.6 &        &      & 0.333   &e \\  
J230306.0$-$312737 & 23  03  06.04 & $-$31  27  37.4 & 18.76 & $-$0.60 &   0.51 &  $<$2.8 &        &      & 2.44?   &  \\  
J230443.6$-$311107 & 23  04  43.60 & $-$31  11  07.5 & 19.49 & $-$0.85 & $-$0.07 &  $<$2.8 &        &      &         &  \\  
J231749.0$-$285350 & 23  17  49.00 & $-$28  53  50.2 & 19.51 & $-$0.70 & $-$0.21 & $<$2.8  &        &      &         &  f \\  
J232531.3$-$313136 & 23  25  31.36 & $-$31  31  36.0 & 19.54 & $-$0.87 &   0.24 & $<$3.0  &        &      &         &  \\  
J234414.7$-$312304 & 23  44  14.70 & $-$31  23  04.3 & 19.65 & $-$0.63 &   0.30 & 5.4  &        &      &         &  \\  
\hline
\end{tabular}
\end{center}
\begin{flushleft}
a 1RXS J014309.8$-$320053\\
b 1RXS J023536.7$-$293845, classification: BL Lac\\
c 1RXS J030416.4$-$283215; this object has an apparent proper
motion of 2.75$\sigma$ (see section 3.2) but is classified as a BL Lac\\
d 1RXS J105534.3$-$012604, classification: BL Lac \\
e objects with follow-up high resolution, high SNR spectra\\
f objects with apparent proper motion between $2-3\sigma$ (see section
2.4)\\ 
g 1RXS J221104.5-284941\\
\end{flushleft}
\end{table*}

Of the 74 objects originally selected, 18 exhibited proper motions
detectable at $\ge3\sigma$.  Of these 18 objects, the typical apparent
proper motions ranged from 30 to 100\,mas\,yr$^{-1}$, largely
consistent with that expected for DC white dwarfs.  One object out of
the 18 exhibited a proper motion of $320\pm 80$\,mas\,yr$^{-1}$, but
this was based on a plate pair with only a one year separation in
epoch.  The distribution of proper motions for our sample is shown in
Fig. \ref{pm}, where we plot the ratio of the proper motion to the
error in each coordinate.  A circle of radius $3\sigma$ then defines
our proper motion cut.  We test whether the distribution is consistent
with the apparent proper motion distribution for all the QSOs in the
2QZ with $18.25<\bj<19.75$ and SNR $\geq10$, shown in
Fig. \ref{qpm}.  A 1-dimensional Kolmogorov Smirnov (KS) test on the
radial distribution of proper motions in Figs. \ref{pm} and \ref{qpm}
 revealed no significant difference at the 95 per cent confidence level
 between the two distributions  when objects exhibiting $>3\sigma$ proper motions are removed
from both samples. For the QSOs, plate measurement errors have resulted
in $\sim$ 2 per cent of sources being
attributed with $> 3\sigma$ proper motion; this is higher than would
be expected for a purely Gaussian distribution. 
  
\begin{table*}
\caption{Featureless continuum objects with proper motions greater
than 3$\sigma$}
\label{table_pm}
\begin{center}
\begin{tabular}{|l|r|r|c|r|r|r|r|} \hline
\multicolumn{6}{|c|} { } &
\multicolumn{2}{|c|} {Proper motion}\\
\multicolumn{1}{|c|} {object} &
\multicolumn{1}{|c|}  {RA (2000)} & 
\multicolumn{1}{|c|} {dec (2000)} &
\multicolumn{1}{|c|} {$\bj$} &
\multicolumn{1}{|c|} {$u-\bj$} &
\multicolumn{1}{|c|} {$\bj-r$} &
\multicolumn {1}{|c|} {RA } &
\multicolumn {1}{|c|} {Dec.} \\
\multicolumn{6}{|c|} { } &
\multicolumn {1}{|c|} {arcsec yr$^{-1}$} &
\multicolumn {1}{|c|} {arcsec yr$^{-1}$} \\
\hline
J002529.8$-$310433 &00 25 29.85  & $-$31 04 33.9  &  19.45  & $-$0.51  &   0.49  &   0.030 $\pm$0.012 & $-$0.039 $\pm$0.011\\
J025425.9$-$315834 &02 54 25.99  & $-$31 58 34.1  &  19.38  & $-$1.01  &   0.08  &   0.035 $\pm$0.013 &   0.031 $\pm$0.012\\
J030459.3$-$274726 &03 04 59.38  & $-$27 47 26.2  &  18.86  & $-$0.46  &   0.52  &   0.042 $\pm$0.010 &   0.002 $\pm$0.010\\
J031435.1$-$310544 &03 14 35.17  & $-$31 05 44.9  &  18.26  & $-$0.82  &   0.02  &   0.030 $\pm$0.010 & $-$0.003 $\pm$0.010\\
J102702.2+003403  &10 27 02.25  &$+$00 34 03.1  &  19.52  & $-$0.89  &   0.34  &   0.323 $\pm$0.083 &   0.123 $\pm$0.083\\
J105037.7$-$011557 &10 50 37.73  & $-$01 15 57.3  &  19.44  & $-$0.89  &   0.20  & $-$0.088 $\pm$0.018 & $-$0.004 $\pm$0.018\\
J113039.1$-$004023 &11 30 39.10  & $-$00 40 23.8  &  19.01  & $-$0.71  & $-$0.08  & $-$0.025 $\pm$0.015 &   0.043 $\pm$0.014\\
J122724.0$-$004317 &12 27 24.05  & $-$00 43 17.4  &  18.68  & $-$0.70  &   0.34  & $-$0.095 $\pm$0.024 & $-$0.013 $\pm$0.026\\
J122809.4$-$005407 &12 28 09.44  & $-$00 54 07.3  &  19.21  & $-$1.48  & $-$0.01  & $-$0.076 $\pm$0.029 & $-$0.073 $\pm$0.031\\
J124034.9$-$014500 &12 40 34.96  & $-$01 45 00.9  &  19.47  & $-$0.76  &   0.18  & $-$0.061 $\pm$0.013 &   0.001 $\pm$0.012\\
J132023.7+001606  &13 20 23.70  &$+$00 16 06.4  &  19.33  & $-$0.97  &   0.29  & $-$0.035 $\pm$0.010 &   0.042 $\pm$0.010\\
J132326.9$-$004157 &13 23 26.96  & $-$00 41 57.7  &  19.44  & $-$0.82  &   0.20  & $-$0.008 $\pm$0.011 & $-$0.054 $\pm$0.010\\
J132849.0$-$002427 &13 28 49.05  & $-$00 24 27.2  &  19.56  & $-$0.71  &   0.51  &   0.022 $\pm$0.011 & $-$0.069 $\pm$0.010\\
J133703.1$-$003910 &13 37 03.16  & $-$00 39 10.8  &  19.08  & $-$0.94  &   0.01  & $-$0.040 $\pm$0.016 &   0.044 $\pm$0.015\\
J142127.9$-$020033 &14 21 27.99  & $-$02 00 33.9  &  18.85  & $-$1.00  & $-$0.01  & $-$0.036 $\pm$0.010 &   0.020 $\pm$0.009\\
J215530.4$-$312148 &21 55 30.49  & $-$31 21 48.6  &  18.70  & $-$0.33  &   0.01  &   0.054 $\pm$0.010 &   0.004 $\pm$0.009\\
J231219.3$-$280928 &23 12 19.36  & $-$28 09 28.1  &  19.08  & $-$0.71  &   0.34  & $-$0.027 $\pm$0.011 & $-$0.045 $\pm$0.010\\
J233139.6$-$272018 &23 31 39.63  & $-$27 20 18.3  &  18.83  & $-$0.77  &   0.19  &   0.035 $\pm$0.011 &   0.087 $\pm$0.010\\

\hline
\end{tabular}
\end{center}
\end{table*}

Rejecting the 18 objects with proper motions from our data set we
define a sample of 56 continuum objects with non-significant proper
motions.  These objects are listed in Table \ref{table_bl}, while we
list the sources with proper motions in Table \ref{table_pm}.  The
sample of 56 objects we refer to below as the 2dF BL Lac (2BL) sample.

\begin{figure}
{\hspace*{0.3in}{\psfig{file=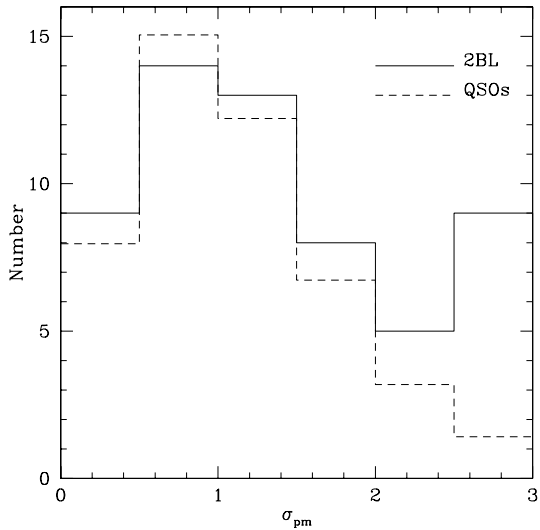,height=3in}}}
\caption{Histogram of proper motions for 2BL sample 
and 2QZ QSOs (re-normalised).}
\label{pmhist}
\end{figure}

However some residual contamination of the 2BL sample is likely to
remain.  This is confirmed by a comparison of the QSO proper motion
histogram with that of the 2BL sample (see Fig.\ \ref{pmhist}).  We
normalised the QSO histogram to contain the same number of objects as
the 2BL sample with $<1\sigma$ proper motions.  There is clearly an
excess in the numbers of objects in the 2BL with $2-3\sigma$ proper
motions.  This excess amounts to approximately 10/56 objects or $\sim
15 - 20$ per cent of the sample.  However, we chose not to enforce a more
stringent proper motion limit on the 2BL, given the observed plate
measurement errors and since two
previously catalogued BL Lacs (J030416.4--283215 and J105534.3-012604) 
independently
`re-discovered' in the 2BL, exhibit  $> 2\sigma$ proper motion
($33\pm12$ and $71\pm35\,$mas$\,$yr$^{-1}$ respectively).  Removing
the nine  radio-quiet objects with
$>2.5\sigma$, or the 12 with $>2.0\sigma$ did not change the
slope of the $n(\bj)$ relation computed for the sample (see section 4.2).

We also found that 35 per cent of the DA white dwarfs from the 2QZ in
the same magnitude range as the 2BL exhibited $> 3\sigma$ proper
motion.  However, we cannot use this number to compute the likely
number of DC white dwarfs which remain undetected in the 2BL due to
low proper motions.  The DA white dwarfs in the 2QZ will have absolute
magnitudes up to $M_B\sim 10$ (Boyle 1989), two magnitudes brighter than
the brightest DC white dwarfs.  Their tangential velocity distribution
is also biased towards lower velocities ($\sim 40-50\,$km$\,$s$^{-1}$)
than non-DA white dwarfs (Sion et al. 1988).  Thus we would expect DA
white dwarfs to exhibit proper motions typically up to three times
less than the DC white dwarfs; $15\,$mas$\,$yr$^{-1}$ for the
brightest $M_B=10$ objects and below the level of detectability with
the plate material available to the 2BL.  For DA white dwarfs with
$M_B=12$, the median magnitude for DA white dwarfs in the Durham/AAT
survey (Boyle 1989), we would predict a mean proper motion of $\sim
40\,$mas$\,$yr$^{-1}$.  Thus the observation that 35 per cent of DA
white dwarfs in the 2QZ have proper motions at the
30$\,$mas$\,$yr$^{-1}$ level or higher is consistent with a population
with a median $M_B=12$ and tangential velocity
$50\pm10\,$km$\,$s$^{-1}$.

By the same token, we anticipate that the vast majority
of the DC white dwarf population with a brightest absolute magnitude of 
$M_B=12$ and somewhat higher observed mean tangential velocity 
($\sim 60$km$\,$s$^{-1}$) will have observed proper motions at greater 
than the $30\,$mas$\,$yr$^{-1}$ level. 
 
We can also obtain an estimate of the catalogue completeness based on
the fraction of previously known BL Lacs in the 2QZ region
re-discovered by this survey.  The Veron \& Veron-Cetty (2001) catalogue
lists 20 objects with a BL Lac or possible BL Lac identification
in the region covered by the 2BL survey.  Seven of these objects
were classified as non-stellar in the initial measurement 
process of the photographic material by the APM and so were not included 
in the stellar catalogue from which the 2QZ was created. 
Only two of these non-stellar objects have magnitudes which fall within the 
magnitude range of the 2QZ.  

As an aside we note that the 2dF Galaxy Redshift Survey independently
re-discovered the
six non-stellar BL Lacs that lie within the magnitude limits
of that survey 
(Lewis I.J., private communication).

Of the remaining 13 objects classified as stellar, 11 have
$u-\bj$/$\bj-r$ colours which place them in the locus used to define
the 2QZ (see Section 4.1, Fig. 9).  The two objects with colours outside this locus are both
classified as BL? in the Veron \& Veron-Cetty catalogue.  Seven of the
11 colour-selected objects lie within the $\bj$ magnitude limits of
the 2QZ and 4 within the magnitude limits and SNR limits of the
2BL.  Three of these objects (1RXS J10555$-$0126, 1RXS
J023536.7$-$293845, and 1RXS J030416.4$-$283215) were independently
re-discovered in the 2BL.  The fourth (UM566), identified as a `?' in the
Veron \& Veron-Cetty BL Lac catalogue, was identified
spectroscopically as an unambiguous DA white dwarf in the 2QZ.

If we restrict our attention to those objects with a definitive BL Lac
classification in the Veron \& Veron-Cetty catalogue, we find that the
2QZ colour criteria would have selected 7/7 known bona fide BL Lacs
with stellar images, or 7/9 BL Lacs if we include the non-stellar BL
Lacs.  Extending this to include the uncertain BL Lacs as well (but
excluding the known white dwarf) gives a figure of 10/12 lying within
the colour critera of the 2QZ.

We therefore conclude that any incompleteness resulting from the 2QZ
colour selection is low (10-15 per cent), and indeed minimal for
previously confirmed BL Lacs. Furthermore all known BL Lacs which made
it through the colour selection process, and fulfilled the magnitude
and SNR criteria of the 2BL, were classified as BL Lacs.

\subsection{Cross-correlation with radio and X-ray surveys}

For each object in the 2BL sample a search was performed using the
NVSS\footnote{NRAO/VLA Sky Survey (Condon et al. 1998)} and (where
applicable) FIRST\footnote{Faint Images of the Radio Sky at 20cm
(White et al. 1997)} databases. Initially nine matches were found in
NVSS with no additional matches in FIRST, and all were within a radius
of 11 arcsec (search conducted out to 15 arcsec).  Subsequent
inspection of all NVSS and FIRST radio maps at the coordinates of our
sources revealed no detections above the rms noise levels,
nevertheless a conservative 
5$\sigma$ detection limit was  
 adopted for the 46 non radio-detected 2BL objects, as
listed in Table 1.  A further cross-correlation of 2BL sources was
carried out with the {\it ROSAT} bright and faint all sky X-ray
catalogues (Voges et al. 1999), initially using a 40 arcsec radius. All five matches found
were, however, within a 15 arcsec radius. A search out to 60 arcsec
revealed no further matches. All five objects with a detectable X-ray
flux are radio sources, consistent with the observation that all X-ray
selected BL Lacs are radio-loud (Stocke et al.\ 1990). Two of these
objects with both radio and X-ray flux are listed in the literature as
BL Lacs (see Table 1).

\section{Radio and Optical Observations}

\subsection{2dF observations}

The 2dF spectra for the 56 objects in the 2BL sample are plotted in
Appendix A.  A number of objects (12/56) also have a second (or
third) observations with 2dF with SNR of least 5.  These are also
shown in Appendix A.  Of the four objects which have two or more SNR
$>10$ spectra, three (J002751.5--293506, J100253.2--001725 and
J114137.1--002730) had both spectra (in one case all three)
independently identified as belonging to a continuum source.  In the
fourth case (J023405.5-301519) one spectrum displays broad lines of
\CIII\ and \CIV\ at $z=1.710$ (see below).  Of the eight objects with
additional lower SNR spectra ($5<$SNR$<10$), two (J003058.2--275629 and
J114327.3--005050) show evidence for broad emission lines.  We have
further examined all the available 2dF spectra to identify weak lines
indicating a redshift, or absorption lines from intervening systems
placing a lower limit on a redshift.  As well as intrinsic redshift
measurements we should also expect that 
if the 2BL sample is largely composed of extragalactic sources a
number should also show narrow metal absorption systems,
e.g. Mg{\small~II} or C{\small~IV}, or indeed Lyman 
$\alpha$ for objects with $z>2.1$.  These can then be used to place
a lower limit on an object's redshift.  In their analysis of the QSOs
in the 2QZ 10k catalogue, Outram et al.\ (2001) find that 11 per cent
of QSOs at $z>0.5$ with SNR $>15$ have significant narrow absorption
line systems with more than one absorption line.  This then suggests
that we should find $\sim 4-5$ sources in the 2BL with these
features.  

Below we discuss the individual sources which show emission or
absorption features:

{\bf J003058.2--275629} has two 2dF spectra, of which
one shows clear evidence of broad \CIII\ and \CIV\ emission lines at
$z=1.577$.  With knowledge of its redshift, it is possible to see weak
\CIII\ ($W_{\lambda}\sim4$\AA $ \ ${\small[rest frame]}) and (tentatively) weak
\CIV\ in the higher SNR spectrum, from which the original
continuum ID was made.  There thus appears to be continuum variability
in this source.

{\bf J023405.5-301519}, as noted above, shows broad
\CIII\ and \CIV\ emission at $z=1.710$ in one spectrum.  There is no
evidence of these features in the other 2QZ spectrum of this object.  We class this a source
with variable continuum emission.

{\bf J031056.9--305901} has a number of narrow absorption lines, most
probably due to intervening absorbers.  The absorber at 4760\AA\ could
be either \MgII\ or \CIV, suggesting a lower limit on the redshift of
$z>0.70$.

{\bf J113900.5--020140} has one strong absorption line visible at
4500\AA\ in both spectra.  A redshift lower limit of $z>0.61$ may be
set assuming this feature is due to the \MgII\ doublet.

{\bf J114221.4--014812} exhibits a single very weak emission line 
($EW_{\rm rest}=4$\AA) at 6370\AA.  Assigning this to \MgII\ we derive a
redshift of $z=1.276$.  However, we see no corresponding
\CIII\ emission at this redshift. 

{\bf J114327.3--005050} has two spectra, of which one shows broad
\CIII\ and \CIV\ emission at $z=1.591$.  The other spectrum shows no
evidence of emission lines.  This object is therefore classed as
variable. 

{\bf J120015.3+000552} shows an absorption doublet at 5400\AA\ which
corresponds to \MgII\ at $z=0.94$.  There is also a weak \FeII\
2600\AA\ absorption line at the same redshift.  The redshift of this
object is thus $z>0.94$.

{\bf J123437.6--012953} contains a system with \MgII,
\FeII\ 2600\AA\ and \FeII\ 2383\AA\ absorption at $z=1.06$, which
defines a lower limit to the redshift of this object.

{\bf J142526.2--011826} shows narrow H$\alpha$ and \OII\ yielding a
low redshift of $z=0.041$.  In this case the \CaII\ H\&K break
contrast is negligible, consistent with the operational definition for
BL Lac classification (Morris et al. 1991).

{\bf J221105.2--284933} contains a strong absorption system at
$z=1.85$ with \CIV , \AlII\ 1671\AA\ and \FeII\ 2344, 2383, 2587 and
2600\AA.  On the blue wing of the \CIV\ absorption line is an emission
feature which could be intrinsic \CIV\ emission, however, no other
emission features are found, and the rest equivalent width of the line
is only 1.6\AA.  We therefore use $z=1.85$ as a lower limit to the
redshift of this object, noting however that this absorption may be
intrinsic to the source.

{\bf J230306.0--312737} shows features bluewards of 4250\AA\ 
which could be Ly$\alpha$ absorption.  There is also a bump at 5320\AA\
which could be associated with \CIV\ at $z=2.44$, however this would
imply that any Ly$\alpha$ emission has been absorbed.  Although this
object may be identified as a QSO at $z=2.44$, we choose to leave it
in the current sample given the uncertainty of its identification.
Removal of this object from the 2BL has no effect on our conclusions.

For the objects which appear to vary we have confirmed that the
featureless nature of the spectra is not an artifact of the
data-reduction process, so that the difference between the two spectra
appears to be real.  We interpret this as due to continuum variablity,
with the broad emission lines appearing when the continuum is in its
low state  - behaviour also shown by the prototype of the BL Lac class
(Corbett et al.\ 2000).  Indeed it is possible that the variable
nature of the continuum in these sources has caused us to miss
potential members of this class when classification is made from a
single spectrum.  Continuum variability would tend to give rise to an
enhanced visibility of emission lines in spectra taken in a low
continuum state -- corresponding to a lower SNR spectrum for a survey
with largely fixed exposure times.

Of the 12 2BL objects for which we have two or more spectra (26
spectra in total), three show some evidence for continuum variability
which would have resulted in a QSO, rather than a continuum
classification.  Given that this behaviour is observed in the
prototype of the class, we choose to retain these objects in the 2BL.
Although based on a small sample (3 spectra out of 26), we estimate
that continuum variability may lead to a potential 10--15 per cent
incompleteness due to mis-classification in the 2BL.

The number of 2BL sources which contain metal absorption lines is
broadly consistent with that expected and we note that a similar
radio-quiet continuum source has been found by Croom et al. (in
preparation) in the 2QZ.  This object falls below the magnitude limit
of the 2BL, but shows a red continuum with \CIV\ and \AlIII\
absorption at $z=1.89$, confirming that such high redshift continuum
sources do exist.  We further note that Fan et al.  (1999) have found a
``quasar without emission lines'' at $z=4.62$ in data from the Sloan
Digital Sky Survey.  
 
\subsection{Higher signal-to-noise spectroscopy}\label{spec_section}

Despite the anticipated difficulty of obtaining redshifts for the 2BL
sample, we nevertheless began a campaign to obtain higher
signal-to-noise, higher resolution spectra of as many of the 2BL
sample as possible.  We can use these observations to place yet more
stringent limits on the featureless nature of the optical spectra and
obtain an independant measure of the likely contamination level of the
2BL by galactic stars.

\begin{figure*}
\centering
\centerline{\psfig{file=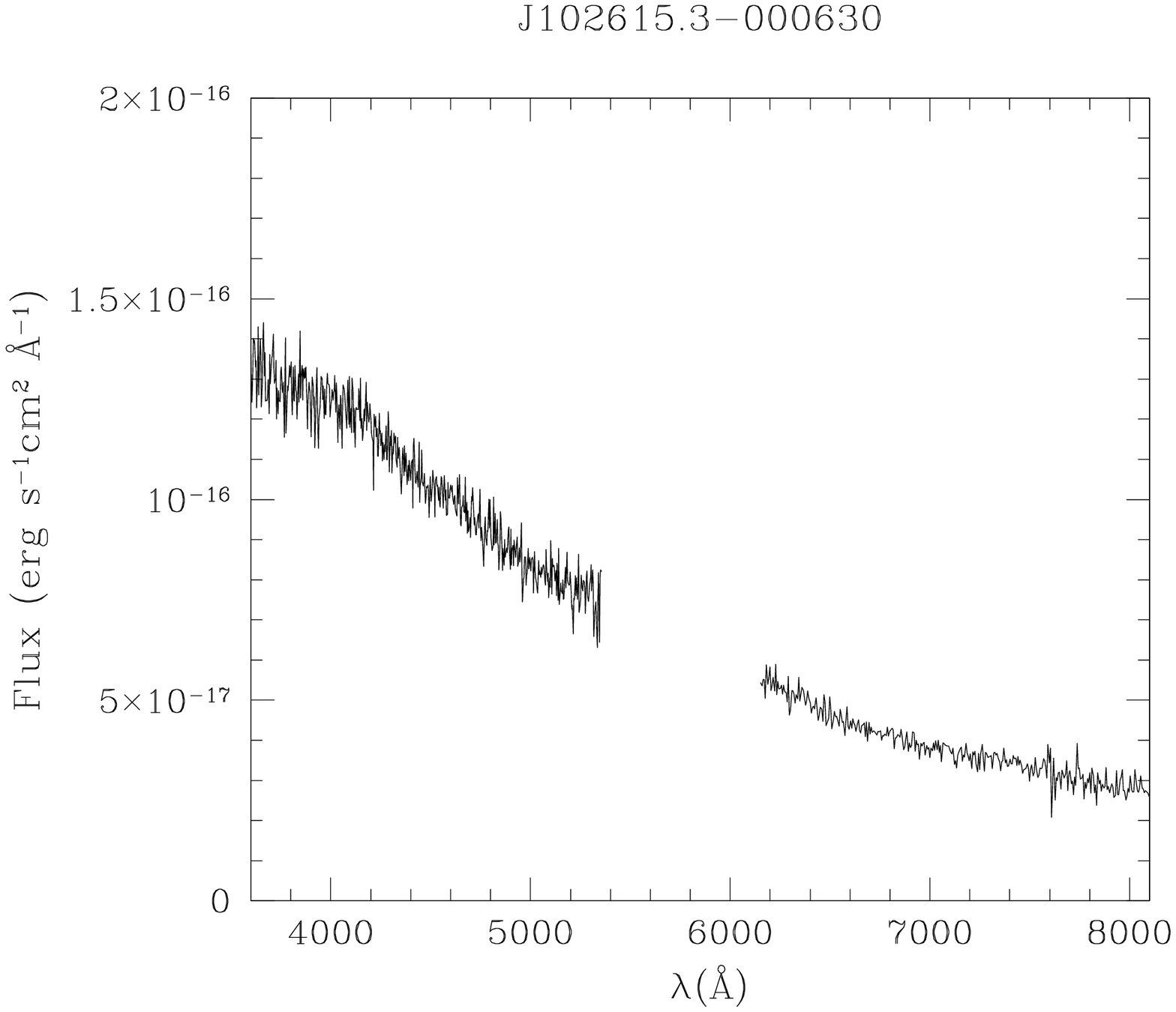,width=7.6cm}\psfig{file=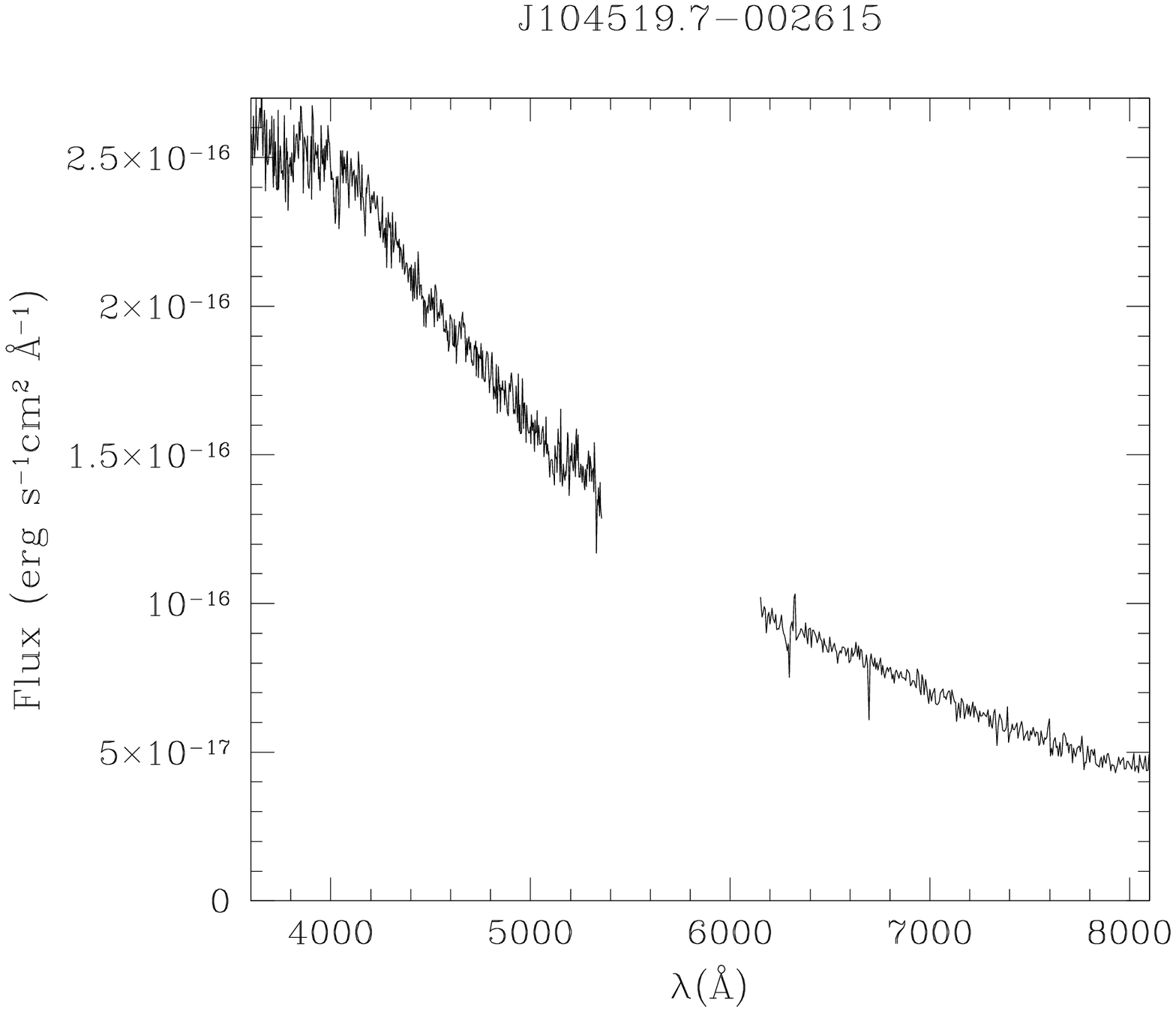,width=7.6cm}}
\centerline{\psfig{file=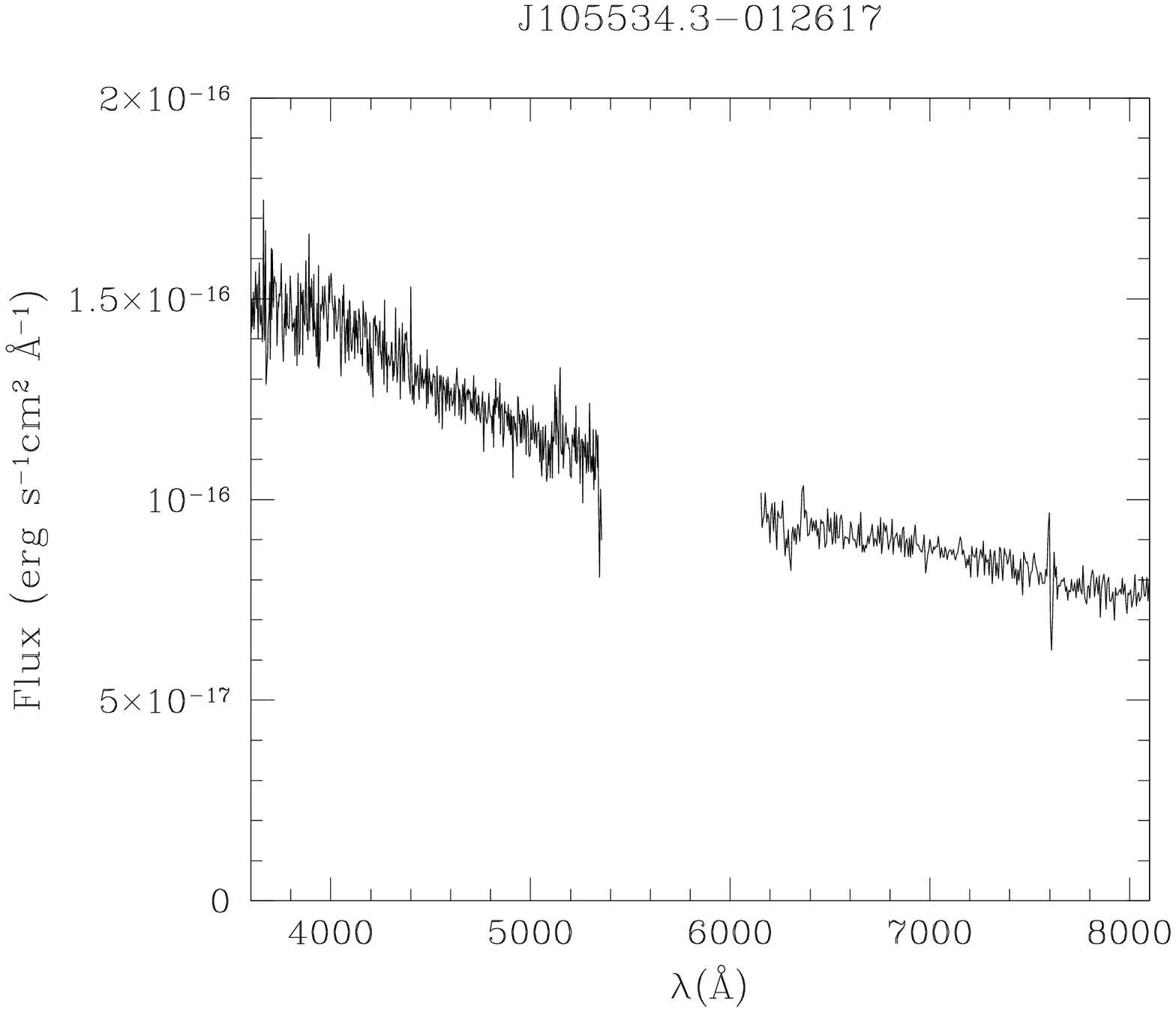,width=7.6cm}\psfig{file=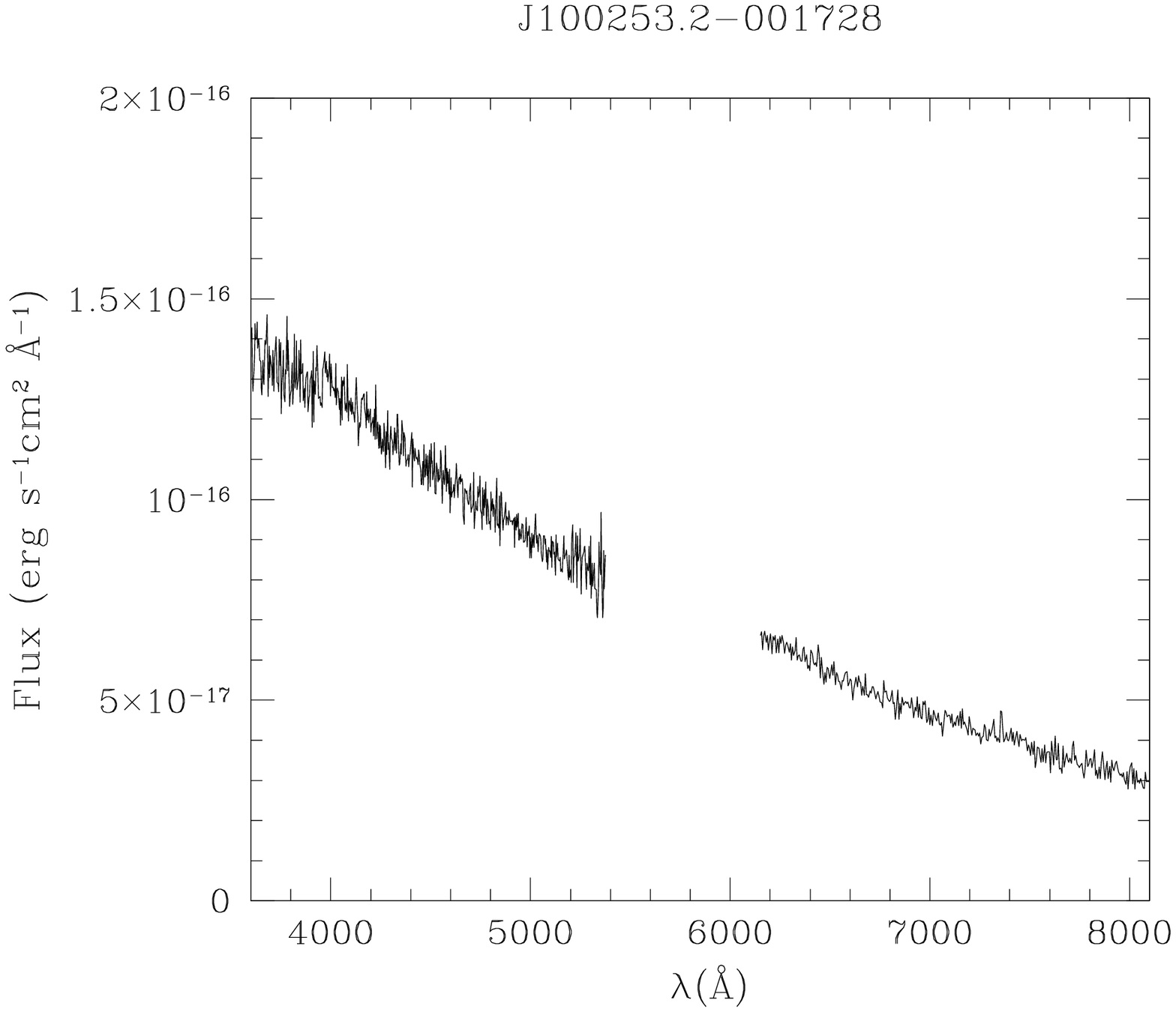,width=7.6cm}}
\centerline{\psfig{file=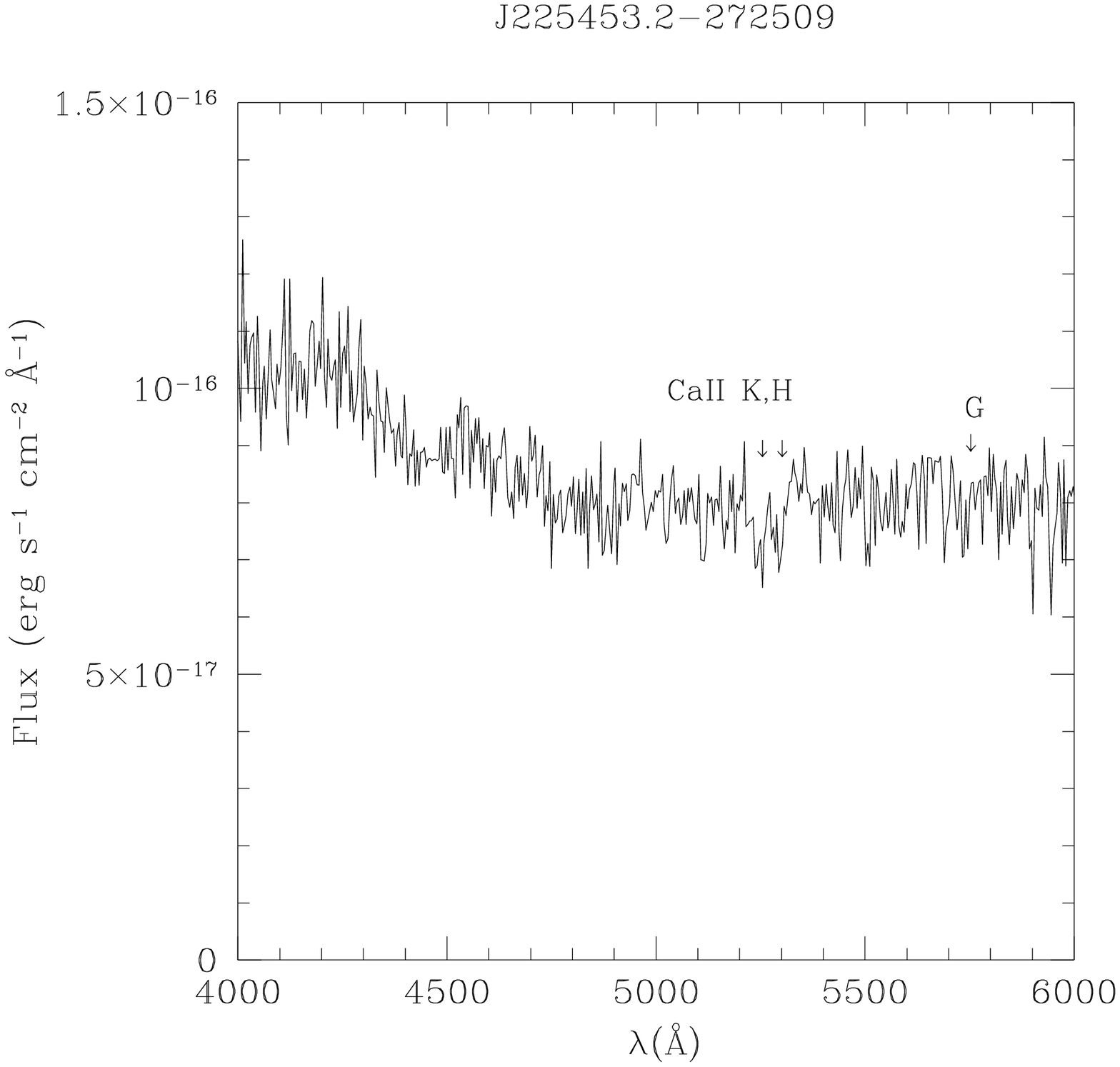,width=7.6cm}}
\caption{Follow up high-resolution, high signal-to-noise spectra for five
 sources in our sample (four WHT and one MSSSO 2.3m spectra).}
\label{spec_fig} 
\end{figure*}

\begin{figure*}
\centering
\centerline{\psfig{file=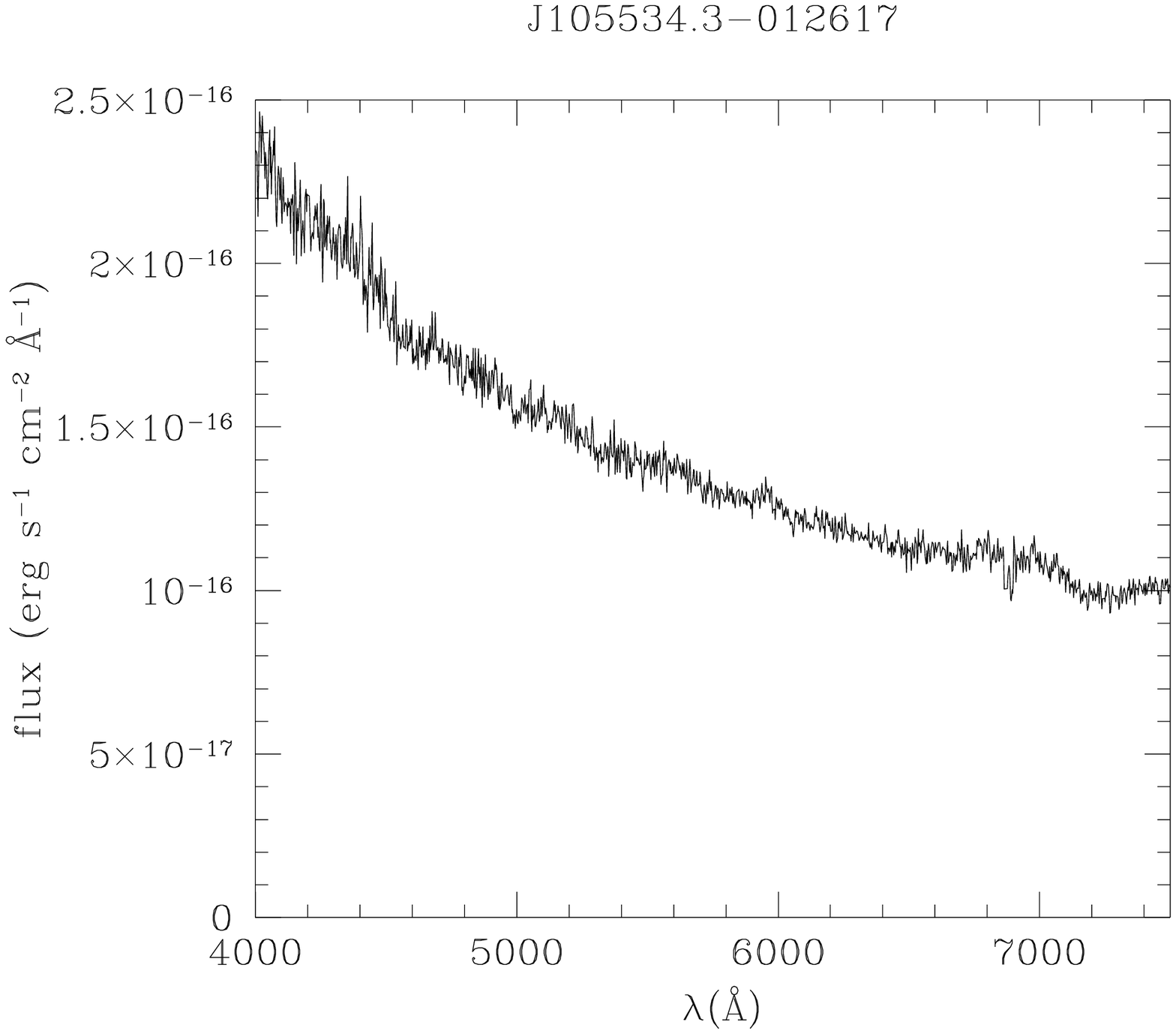,width=8.0cm}\psfig{file=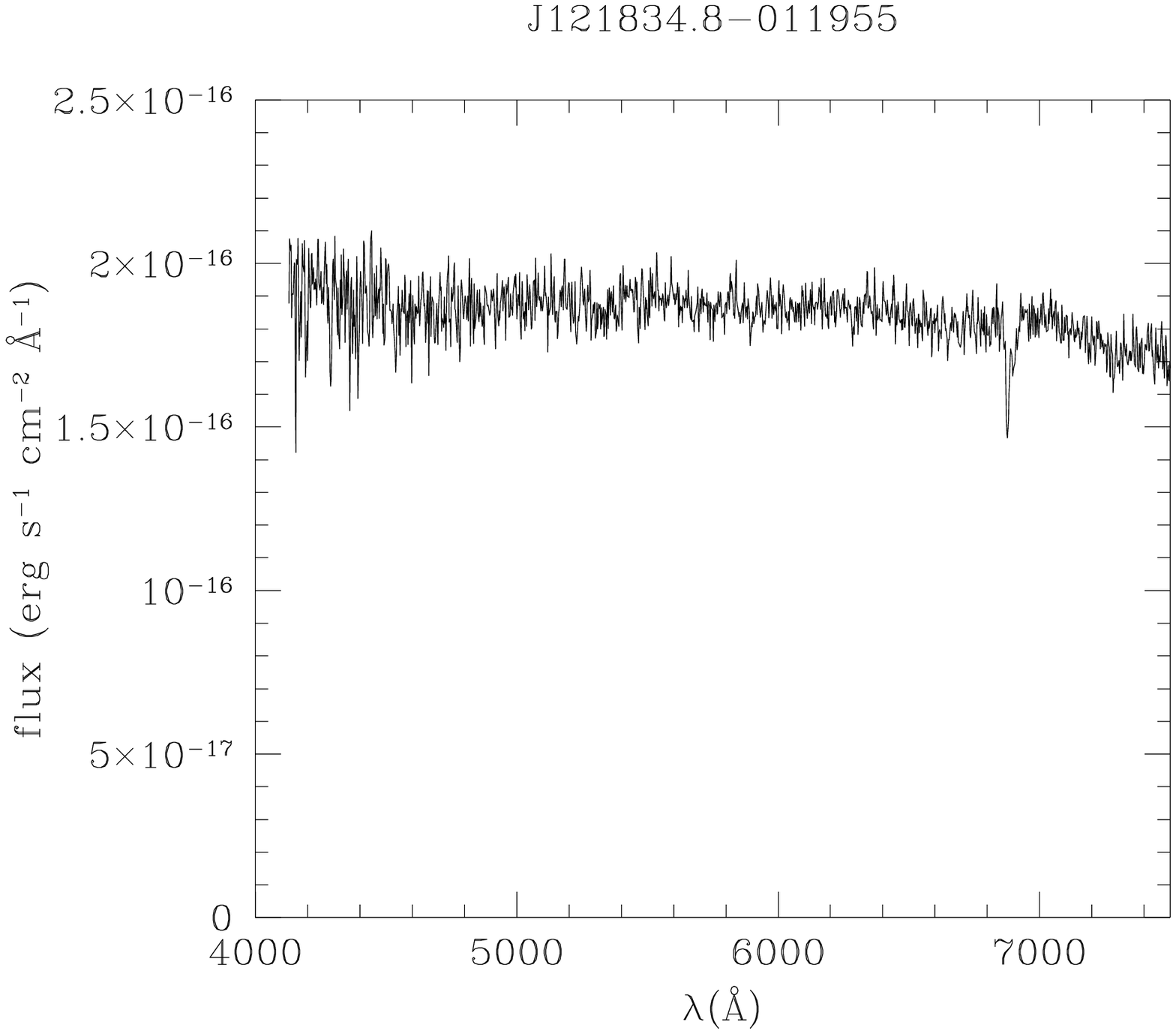,width=8.0cm}}
\caption{The two radio loud objects observed with Keck by Brotherton
et al. 1998. The feature at $\sim$ 6900\AA $ \ $ is a telluric absorption
band.}
\end{figure*}

One candidate BL Lac (J225453.2--272509) was observed in August 2000
using the long slit, double-beam spectrograph (DBS) on the Mt Stromlo
2.3m telescope at Siding Spring. The object was observed for a total
of two hours with resolutions of 4.1 and 4.5 \AA\ pixel$^{-1}$ in the
blue (3800--6100\AA) and the red (6200--8500\AA) arms of the DBS
respectively.  Data reduction was performed using standard
IRAF\footnote{IRAF is distributed by the National Optical Astronomy
Observatories,operated by the Association of Universities for Research
in Astronomy Inc., under cooperative agreemfent with the National
Science Foundation} procedures ({\it ccdred} and {\it apall}).\

J225453.2--272509 was confirmed as having no emission lines and we
were able to measure a redshift of $z=0.333$ from \CaII\ H \& K and G
band absorption features.

A further 4 BL Lac candidates were observed on 2001 February 21--22
using the ISIS double-beam spectrograph on the William Herschel
Telescope. These spectra covered the range 3800\AA --5000\AA $\ $ and
6300\AA--8000\AA\ in the blue and red arms respectively at a spectral
resolution of 1\AA$\,$pixel$^{-1}$.  Exposure times of 20 minutes were
sufficient to yield a SNR $>$25 in the continuum of these objects.

Two candidates in the 2BL sample (J105534.3--012617 and
J121834.8--011955) had also been observed with Keck as part of a
follow-up of radio-loud 2QZ sources (Brotherton et al.\ 1998). The
first of these two was one of the sources also observed with the WHT.

These 7 objects represent 12.5 per cent of the 2BL sample.  Apart from
the Keck spectra, the objects on which follow-up spectroscopy was
obtained were selected at random from the 2BL.  Of the five objects
selected at random, all exhibit spectra consistent with a continuum
source identification.

We are thus able to obtain possible redshifts/limits for 12 objects in this
sample, covering a relatively wide range in redshift; from $z=0.041$
to $z>2.4$, with a median $z=1.2$ (assuming that objects with
lower limits are at that limiting redshift).  Of these, only two are
detected radio sources (J225453.2--272509, $z=0.333$ and
J221105.2--284933 at $z>1.85$).  

From the optical spectroscopic observations and the proper motion
study above we can place a limit on the likely contamimation of  the
sample by weak-lined Galactic subdwarfs and white dwarfs of  10 -- 20
per cent.  Although both are subject to large uncertainties, we note
that this limit on the contamination level  is similar to the
potential incompleteness level caused by  continuum variability (see
above).  In the statistical analysis below, we therefore choose not to
correct the numbers in the 2BL for either incompleteness or
contamination. However we do note that the numbers/normalisation of
the 2BL sample carry an uncertainty at least at the $\pm 20$ per cent
level.  

\subsection{Radio observations at 8.4 GHz}

Observations of a sub-sample of eight 2BL objects were made at a frequency of
8.4 GHz  using the NRAO Very Large Array on 2001 January
28.  The array was in a hybrid BnA$\rightarrow$B configuration with 25
antennas operating.  BL Lac candidates from the equatorial strip were
observed for 32 minutes each while those from the southern strip
($-27^o < \delta < -31.5$) were observed for 40 minutes, in both cases
in snapshot mode.  Flux densities were bootstrapped from 3C286.  The
typical noise level in each snapshot was 45$\mu$Jy.
\begin{figure}
{\hspace*{0.4cm}{\psfig{file=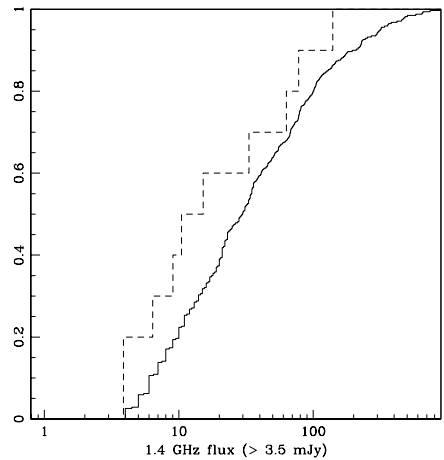,height=6.5cm}}}
\caption{Cumulative distribution of the 340 radio loud QSOs and the 9 BL Lac candidates (dashed line) with $S_{1.4GHz} >$ 3.5mJy}
\label{radio}
\end{figure}
\begin{figure}
{\hspace*{0.4cm}{\psfig{file=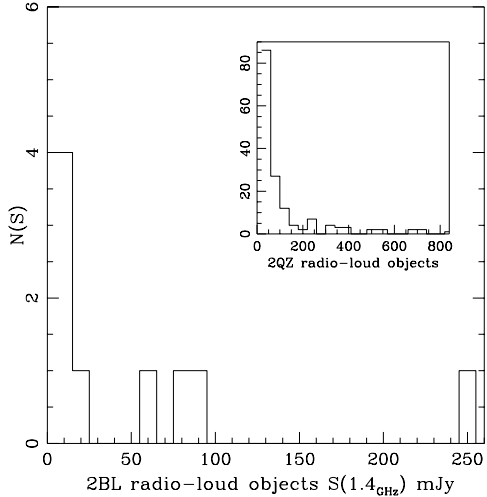,height=3in}}}
\caption{Differential distribution of the 340 radio loud QSOs and the 9 BL Lac candidates}
\label{radio1}
\end{figure}
Data were reduced using the ATNF's\footnote{Australia Telescope
National Facility} \textsc{miriad} software package (Sault, Teuben \& 
Wright 1995), following standard procedues from the Miriad User's
Manual (Sault and Killeen 1999).

We found  no detections at the positions of the candidate
BL Lac sources with a flux density $>5\sigma$ ($\simeq 0.2$mJy) above the background
noise level.  Results are included in
Table 1.\

A comparison of the radio flux distributions of the nine radio-loud
objects in the 2BL and the 340 radio-loud QSOs with $\bj <20.00$ and
$S_{1.4} >$ 3.5mJy in the 2QZ is shown in Figs. \ref{radio} and
\ref{radio1}.  A KS test revealed no significant difference between the
two populations.

For the one object in the 2BL sample with a measured redshift and
documented NVSS flux density of 52.6mJy, we obtain a radio power $P_{1.4}=
2.5\times 10^{25}\,$W Hz$^{-1}$, and $M_B = -22.05 $ (using
$\Omega_{\rm m}=1$, $\Omega_{\rm \Lambda}=0$, $H_0
=50\,$km$\,$s$^{-1}$Mpc$^{-1}$, $\alpha_{\rm r}=-0.5,\ \alpha_{\rm
opt}=-0.5$).  This value lies at the upper end of the range of 1.4 GHz
radio powers calculated for the EMSS BL Lacs (Rector et al.\ 2000)
which range from 4 $\times 10^{23}\,$W Hz$^{-1}$ to 3 $\times
10^{25}\,$W Hz$^{-1}$ using the same cosmology.

\section{Basic properties of the 2BL}

\subsection{Colour distribution}\label{c_distrib}

\begin{figure}
\psfig{file=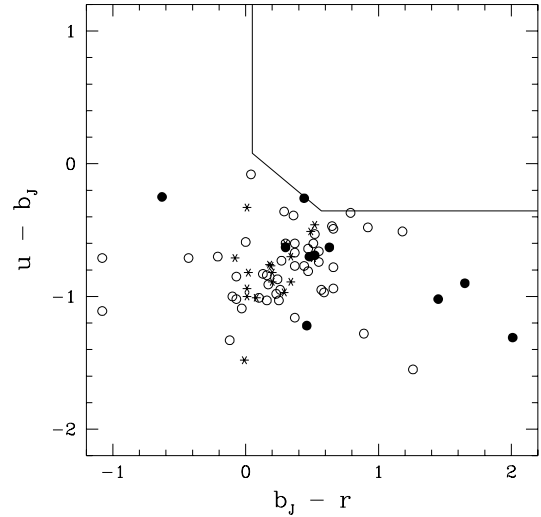,height=3.1in}
\caption{$u-\bj$, $\bj-r$ plot of BL Lac candidates; filled circles
represent 2BL objects with radio detections (1.4 {\sc gh}z) while open circles represent radio-quiet
objects. Asterisks are the 18 featureless continuum objects with
proper motion which have been removed from the 2BL sample.  The straight
lines are the 2QZ colour selection cut-off.}
\label{colour}
\end{figure}

In Fig. \ref{colour} we show the $u$-$\bj$\footnote{u=U-0.24 (Smith et
al. 2001)}, $\bj$-$r$ colour distribution of 2BL
sources. Open circles represent candidate BL Lacs with no radio
detection, while filled circles denote those objects with a measured
radio flux density ($S_{1.4}>1.5$mJy); asterisks represent the 18
objects which were found to have significant proper motion and
subsequently removed from the sample.   A 2-D KS test on the
$u-\bj$/$\bj-r$ distributions of these 18 objects and the 56 objects
in the 2BL sample returned a 95 per cent probability that the two
distributions are different; this increased to 97 per cent if only
those 2BL objects with $< 2\sigma$ proper motions were considered.

\begin{figure}
\psfig{file=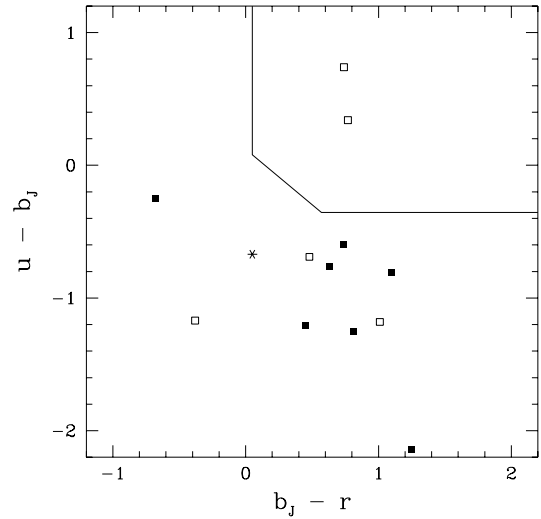,height=3.1in}
\caption{$u-\bj$, $\bj-r$ plot of the 13 objects from the Veron \&
Veron-Cetty (2001) catalogue included in the 2QZ input catalogue. 
Solid squares represent a BL classification while open squares
refer to BL? objects.  The
mis-identified white dwarf is shown as an asterisk.}
\label{vv}
\end{figure}

As a comparison, the $u-\bj$/$\bj-r$ distribution of the 13 BL/BL?
objects from the Veron \& Veron-Cetty catalogue discussed in section
2.4 is plotted in Fig. \ref{vv}.

\begin{figure}
\psfig{file=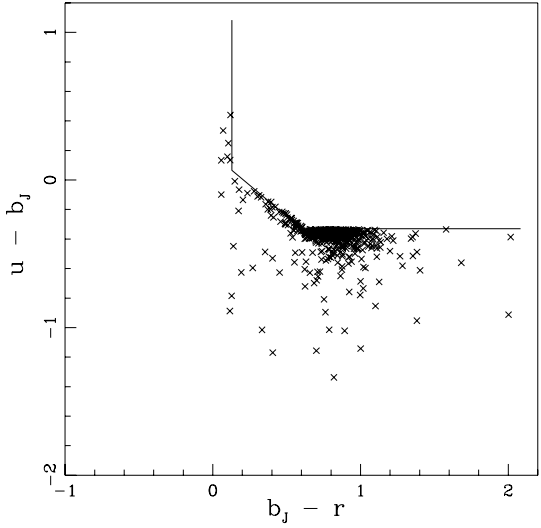,height=2.95in}
\caption{$u-\bj$, $\bj-r$ diagram of F and G type stars in the 2QZ
with SNR $>$ 10.} 
\label{subd}
\end{figure}

\begin{figure}
\psfig{file=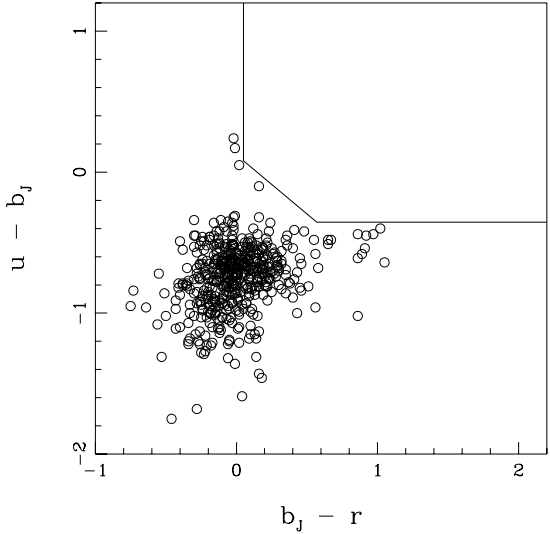,height=3.1in}
\caption{$u-\bj$, $\bj-r$ diagram of DA white dwarfs in the 2QZ
with SNR $>$ 10.} 
\label{da}
\end{figure}

Three of the four strong radio sources ($S_{1.4GHz} > 50$ mJy) show
evidence of an enhanced red component while still exhibiting a strong
UV flux.  A comparison with the colours of Galactic sub dwarfs
(Fig. \ref{subd}) shows that the 2BL is unlikely to be heavily contaminated by
this population. Less than 10 per cent of the 2BL sample exhibit
colours in the densely populated locus of the Galactic subdwarfs. The
region populated by the DA white dwarfs identified in the 2QZ (see
Croom et al.\ 2001) is also shown (Fig. \ref{da}). For a given
$u-\bj$, the mean $\bj -r $ colour for 2BL objects is approximately
0.4 mag redder than that of the DA white dwarfs.  A 2-D KS test on
colour distributions of the 2BL objects and the 2QZ DA white dwarfs
returned a $>$ 99.9 per cent probability
that the two distributions are different.  A similar 2-D KS test was
carried out on the 56 2BL objects and the 10 BL/BL? objects from the
Veron \& Veron-Cetty catalogue that satisfy the
2QZ colour selection criteria (Fig. 9), but excluding the
white dwarf.  The test revealed no significant
difference betwen the two distributions.


\subsection{Optical $n(\bj)$ relation}

\begin{figure}
{\hspace*{0.2cm}{\psfig{file=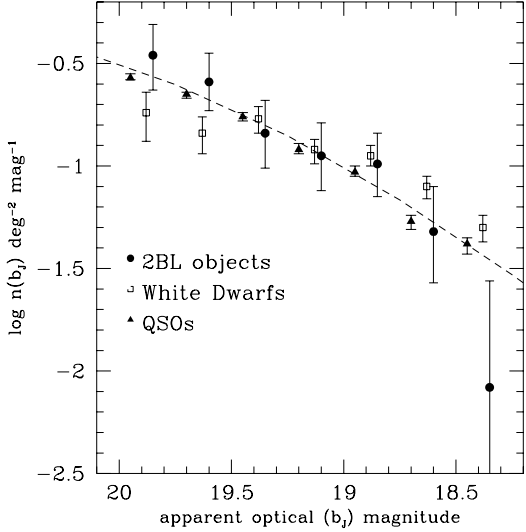,height=3in}}}
\caption{The $n(\bj)$ relation for 2dF QSOs normalised by 0.014
(triangles) and 2QZ white dwarfs (open squares), normalised by 0.12,
plotted against the 2BL sample and the predicted $n(\bj)$ relation
from Boyle et al.\ (2000) (dashed line). The choice of normalising
factor was made in order to minimize the $\chi^2$ statistic between
the 2 distributions. For clarity 2BL objects are offset by 0.03 mag
relative to the white dwarfs. The QSO number counts are from Boyle et
al.\ (2000) }
\label{wd}
\end{figure}

The differential number magnitude, $n(\bj)$, relation for the 2BL is
plotted in Figure \ref{wd}.  The $n(\bj)$ relation is normalised to
the fractional number of objects with signal-to-noise ratio greater
than 10 in each $0.25\,$mag bin:
\begin{equation}
n(\bj)=\frac{N_{\rm obs}(\bj)}{544 \ {\rm deg^2}} \times \frac{N_{\rm TOT}(\bj)}{N_{\rm {\sc snr \geq 10}}(\bj)}
\end{equation}
\noindent as set out in Table 3. 

\begin{table}
\caption{Observed and computed number counts; $N_{TOT}$ refers to the
number of 2QZ objects in the respective magnitude range, and
$N_{{\sc snr}\geq 10}$ to those which also satisfy the criterion of SNR
 $\geq$10. $N_{obs}$ is the number of featureless continuum objects
(i.e. BL Lac candidates) identified from the SNR $\geq$10 spectra,
and $n(\bj)$ is computed as given in equation (1). The errors are 
1$\sigma$ Poisson errors (Gehrels 1986).  }
\begin{center}
\begin{tabular}{|c|c|c|r|c|} \hline
\multicolumn{1}{|c|} {$0.25\,$mag bin } &
\multicolumn{1}{|c|} { $ N_{\sc tot} $}  &
\multicolumn{1}{|c|} { $ N_{\sc snr \geq 10} $ } &
\multicolumn{1}{|c|} {$ N_{obs} $ } &
\multicolumn{1}{|c|} {log $n(\bj)$ } \\
\hline

18.25$<\bj \leq$18.50 &  936 &  819 &  1 
$_{-0.83}^{+2.30}$ & --2.08\\  
  & & & &  \\
18.50$<\bj \leq$18.75 & 1091 &  844 &  5
$_{-2.16}^{+3.38}$ & --1.32\\ 
   & & & &  \\
18.75$<\bj \leq$19.00 & 1368 &  988 & 10 
$_{-3.11}^{+4.27}$ & --0.99\\ 
  & & & &  \\
19.00$<\bj \leq$19.25 & 1759 & 1040 &  9 
$_{-2.94}^{+4.11}$ & --0.95\\  
  & & & &  \\
19.25$<\bj \leq$19.50 & 2433 & 1116 &  9 
$_{-2.94}^{+4.11}$\ & --0.84\\  
  & & & &  \\
19.50$<\bj \leq$19.75 & 3255 & 1106 & 12 
$_{-3.41}^{+4.56}$ & --0.59\\ 
  & & & &  \\
19.75$<\bj \leq$20.00 & 4012 &  859 & 10  $_{-3.11}^{+4.27}$ & --0.46\\ 

\hline
\end{tabular}
\end{center}
\end{table}

Based on a weighted least squares fit of the 2BL $n(\bj)$ we measured
a slope of $n\propto 10^{0.66\pm 0.10m}$.  In order to check the
robustness of the $n(\bj)$ relation against choice of signal-to-noise
limit, we derived the $n(\bj)$ relation for the sample of 27 candidate
2BL objects with SNR $>15$. The best fit displayed an almost identical
slope and normalization, $n(\bj) \propto 10^{0.71 \pm 0.24m}$, to that
obtained for the SNR $ \geq 10$ sample.  The steepness of the $n(\bj)$
relation is also robust against choice of survey magnitude limits; a
least squares fit to only those points with $\bj <$ 19.5 yielded a
slope of 0.64 $\pm$ 0.25. Rejecting the 12 radio-quiet 2BL objects with proper
motions of $2\sigma-3\sigma$ produced a slope of 0.68 $\pm 0.09$,
hence almost identical to that of the full sample of 56 objects.

The $n(\bj)$ relation for all DA white dwarfs with SNR$>15$ in the 2QZ
is plotted in Fig. \ref{wd}.  This SNR was chosen to ensure minimal
contamination of the white dwarf sample by subdwarfs which can be
increasingly confused with white dwarfs in the 2QZ sample for SNR$<15$
(Vennes et al., in preparation).  The white dwarf $n(\bj)$ relation
exhibits a slope of $n\propto 10^{0.37\pm 0.07m}$, significantly
flatter than the $n\propto 10^{0.66\pm0.10m}$ slope of the 2BL
objects.  Indeed, a minimum $\chi^2$ fit between the $n(\bj)$
relations for the 2QZ white dwarf and 2BL samples reveals that they
are different at the 99 per cent confidence level.

\begin{table}
\begin{center}
\begin{tabular}{|l|r|c|} \hline
\multicolumn{1}{|c|} {Sample} &
\multicolumn{1}{|c|} {slope} &
\multicolumn{1}{|c|} {$\sigma$}\\
\hline
2BL & 0.66 & 0.10\\
2QZ QSOs & 0.59 & 0.03\\
2QZ WDs & 0.37 & 0.07\\
\hline
\end{tabular}
\end{center}
\caption{Comparison of $n(\bj)$ values, $n\propto 10^{slope*m}$, for the
2BL objects, the 2QZ white dwarfs and QSOs in Fig.
\ref{wd}.}
\end{table}
We also plot in figure \ref{wd} the re-normalised $n(\bj)$ relation for
the 2QZ QSOs based on a minimum $\chi^2$ fit to the 2BL counts.  The
$\chi^2$ test confirms that the 2BL and re-normalised 2QZ $n(\bj)$
relations are consistent with each other, with a relative
normalisation of 0.014. 

\subsection{Predicted redshift distribution}

Given the similarity between the $n(\bj)$ relations for the 2BL sample
and 2QZ QSOs, we used the evolutionary model derived for the latter to
obtain a prediction of the redshift distribution for the 2BL. In doing
this we assume that the vast majority of 2BL objects represent a
sample of AGN with very weak or absent emission lines and hence
exhibit the same evolutionary behaviour (Croom et al.\ 2001) and
identical form of luminosity function (LF) to that obtained for the
2QZ QSOs (see Boyle et al.\ 2000). To give the observed total number
objects in the 2BL we used a normalisation 0.015 times (consistent
with the $\chi^2$ fitting above of 0.014) that of the QSO LF
(i.e. $\Phi^*_{{\sc 2bl}} =
1.65 \times 10^{-8}$mag$^{-1}$Mpc$^{-3}$), with an identical break
luminosity, $M^*_{(z=0)}=-21.9$.  A $\chi^2$ test confirmed that the predicted
$n(\bj)$ relation based on this model (also plotted in Fig. \ref{wd}) is
consistent with the observed 2BL $n(\bj)$ relation.

\begin{figure}
{\hspace*{0.1in}{\psfig{file=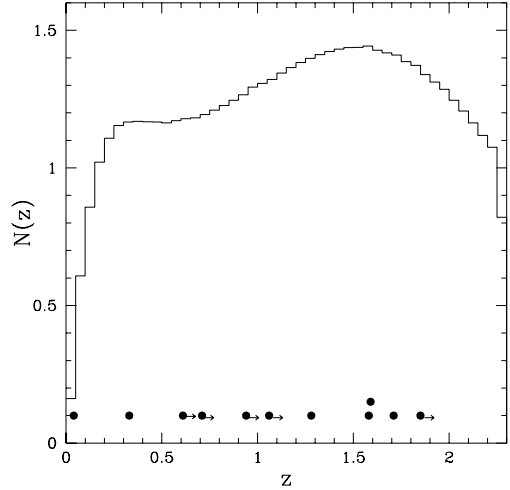,height=2.75in}}}
\caption{Predicted redshift distribution of 2dF BL Lac candidates using the
evolution model of Boyle et al. (2000); dots and dots with arrows indicate the
actual measured redshifts/lower limits for 2BL objects.}
\label{nz}
\end{figure}
The predicted number-redshift, $n(z)$, relation for the 2BL sample
based on this model is shown in Fig. \ref{nz}.  In common with the
QSOs at similar magnitudes ($\bj <20$), the $n(z)$ relation is predicted
to peak
at $z=1.6$ with a median redshift of $z=1.1$.  The decline in the
$n(z)$ at higher redshifts results from a combination of factors: the
depth of the sample, the slow-down in the QSO evolution rate at $z>2$
and the k-correction in the $B$-passband due to the Lyman $\alpha$
forest. The bimodal nature of the $n(z)$ relation is an artefact of
the strong magnitude dependence of the areal coverage function in the
2BL.

If the extragalactic objects in the 2BL sample are distributed with
this $n(z)$ relation it will be hard to obtain redshifts/unambiguous
identifications for these objects from their optical spectra alone
unless they exhibit intervening absorption line systems.
Previously, many BL Lac redshifts have been based on spectroscopic
features in the host galaxy (e.g Ca{\sc II} H\&K, G band).  The 2dF
spectra on which the selection of the sample was based cover the
region 3800--7800\AA\ (with the region redward of 6800\AA\
increasingly contaminated by night sky emission).  If one requires
both Ca{\sc II} H\&K and the G band to be detected for a secure
identification/redshift, then this would only be possible in the 2dF
spectra for objects with $z<0.5$, less than 10 per cent of sample
predicted by the $n(z)$ relation.

Nevertheless, the small number of tentative redshifts or lower limits
on redshifts obtained for objects in the 2BL to date is certainly not
inconsistent with the proposed redshift distribution (see Fig.\ref{nz}).

\section{Discussion}

The 2BL sample is a unique, optically selected sample of featureless
continuum objects. Follow-up optical spectroscopy at higher resolution
and SNR confirms the featureless nature of a subsample of these objects, 
while the broadband optical colours of these objects are
inconsistent with Galactic subdwarfs. The steep $n(\bj)$ relation of
this population implies strong cosmlogical evolution, consistent with
that of QSOs and indicative of a population that is extragalactic in
origin.  Based on a QSO evolution model the median redshift of the
sample is predicted to be $z\sim 1.1$. In all respects, other than their radio
properties, the majority of the 2BL objects are consistent with the
class of objects known as BL Lacs.
 
Hitherto there has been no large optically-selected sample of BL
Lacs. Previously all BL Lacs have been initially identified by their
radio or X-ray emission.  Moreover, all the BL Lacs identified to date
have shown strong radio emission.  Indeed Stocke et al.\ (1990) have
argued that there are no radio-quiet BL Lacs on the basis that all BL
Lacs identified in the EMSS had radio flux densities of $S_{\rm 5GHz} >
1\,$mJy.

Notably all five 2BL objects with X-ray detections (based on the
$ROSAT$ All Sky Survey, RASS) are also milli-Jansky radio sources.
The flux limit of the RASS ($S_{\rm 0.1-2.4keV} = 8 \times
10^{-14}$\,erg\,s$^{-1}$\,cm$^{-2}$) is equivalent to $S_{\rm
0.3-3.5keV} = 6.2 \times 10^{-14}$\,erg\,s$^{-1}$\,cm$^{-2}$ for an
object with an X-ray spectrum of $f_\nu\propto\nu^{-1}$, hence almost
identical to the EMSS flux limit.

\begin{figure}
\psfig{file=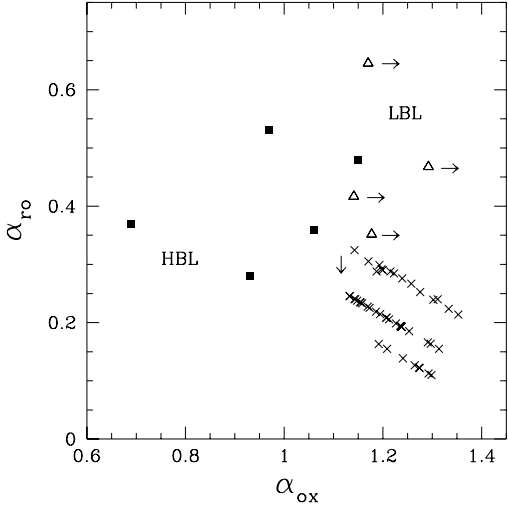,height=3.0in}
\caption{$\alpha_{\rm ro}-\alpha_{\rm ox}$ diagram (1.4 GHz, 4400 \AA, 2Kev) of the 2BL sample.
Five sources (solid squares) have both X-ray and radio flux; open
triangles denote sources with a radio detection and right arrows denote upper
limits of $f_x =$ 8 $\times 10^{-14}$ erg s$^{-1}$ cm$^{-2}$; in
addition crosses have a limiting radio flux density as per Table 1;
$S_{8.4}$ limiting fluxes of 0.2
mJy, have been converted to a 1.4GHz flux density of 0.5 mJy using
$\alpha_r = -0.5$ }
\label{alf}
\end{figure}
 
In Fig. \ref{alf} we plot the X-ray-to-optical $\alpha_{\rm ox}$ and
radio-to-optical $\alpha_{\rm ro}$ flux ratio for the sample, where we
have defined $\alpha_{\rm ro}$ as [log($S_{1.4} / f_{\rm
4400\AA}$)]/5.7 and $\alpha_{\rm ox}$ as [log($f_{\rm 4400\AA} /
f_{2keV}$)]/2.85.  In the absence of redshift information
k-corrections have not been applied, however with $\alpha_{opt} =
\alpha_r = -0.5$ and $\alpha_x = -1$ corrections at $z =$ 1.1 amount
to less than $-0.05$ for $\alpha_{ox}$. In common with Stocke et al.\
(1990) we find that our bright X-ray detected BL Lacs exhibit a narrow
range in radio--optical spectral index, namely $0.28< \alpha_{\rm ro} <
0.53$.  

At the predicted median redshift of the 2BL ($z=1.1$), the $5\sigma$
 0.2 mJy detection limit of the VLA observations corresponds to an
upper limit for the radio power of $P_{\rm 1.4GHz} \leq 1.7 \times 
10^{24}\,$W$\,$Hz$^{-1}$. This is still within the range of radio
powers for the EMSS BL Lac sample observed by Rector et al.\ (2000).
Thus high redshift BL Lacs with radio powers similar to those low
redshift BL Lacs found by Rector et al.\ (2000) could still have
evaded detection at deepest radio flux limits probed by our VLA
observations.  However, these high redshift objects would have a lower
$\alpha_{\rm ro}$ than BL Lac objects observed to date, and would
constitute a distinct population, occupying a different region of the
$\alpha_{\rm ro}-\alpha_{\rm ox}$ plane to that of hitherto known BL
Lac objects. The continuum emission mechanism that would produce such
radio-weak objects is unclear;  these objects may simply be a class 
of lineless QSOs with an optical continuum mechanism (e.g. hot accretion disk)
very different to that producing the optical radiation in objects
hitherto
identified as BL Lacs. 

While the nature of the 2BL objects with no detectable radio emission
remains uncertain, it appears that those objects with radio detections
are bona fide BL Lacs, and are some of the optically faintest BL Lacs
detected to date. In addition, of the 11 or 12 2BL objects that have
redshift information (and hence appear to be extragalactic), 10 are at a redshift greater than 0.6, a region generally
thought to be sparsely populated by BL Lac objects.

If the radio-quiet objects are not AGN, then the question remains as
to their nature.  If they are DC white dwarfs, or indeed any
class of Galactic star, then there is perhaps an even greater puzzle
in explaining the steep $n(\bj)$ relation.  Further observations of the
2BL sample including infrared photometry and spectroscopy (to attempt
to detect the host galaxy) and/or variability and polarisation studies
may help to resolve their identity.

\section*{Acknowledgements}
The 2QZ is based on observations made with the Anglo-Australian
Telescope and the UK Schmidt Telescope; we would like to thank our
colleagues on the 2dF galaxy redshift survey team and all the staff at
the AAT who have helped to make this survey possible. We thank the
anonymous referee for a careful scrutiny of the manuscript and his/her
insightful comments which have greatly improved the contents and
clarity of the paper.  DL thanks the School of Physics at the
University of Sydney for a Postgraduate (Mature Age) Scholarship.


\newpage
$ \ $
\newpage
\appendix
\section{2dF spectra}
\begin{figure*}
\caption{2dF spectra for all the objects in the 2BL sample, including
repeat observations.  Objects with repeat observations are indicated
by an a, b or c after the object name, in descending order of SNR.
The SNR of each spectrum and the object's $\bj$ magnitude is also
shown.  If an object has a measured redshift (not including lower
limits), this is shown together with the positions of spectral
features.  A circled cross marks the location of telluric sky
absorption bands. Straight lines in the spectra indicate the removal of
night sky lines}
\psfig{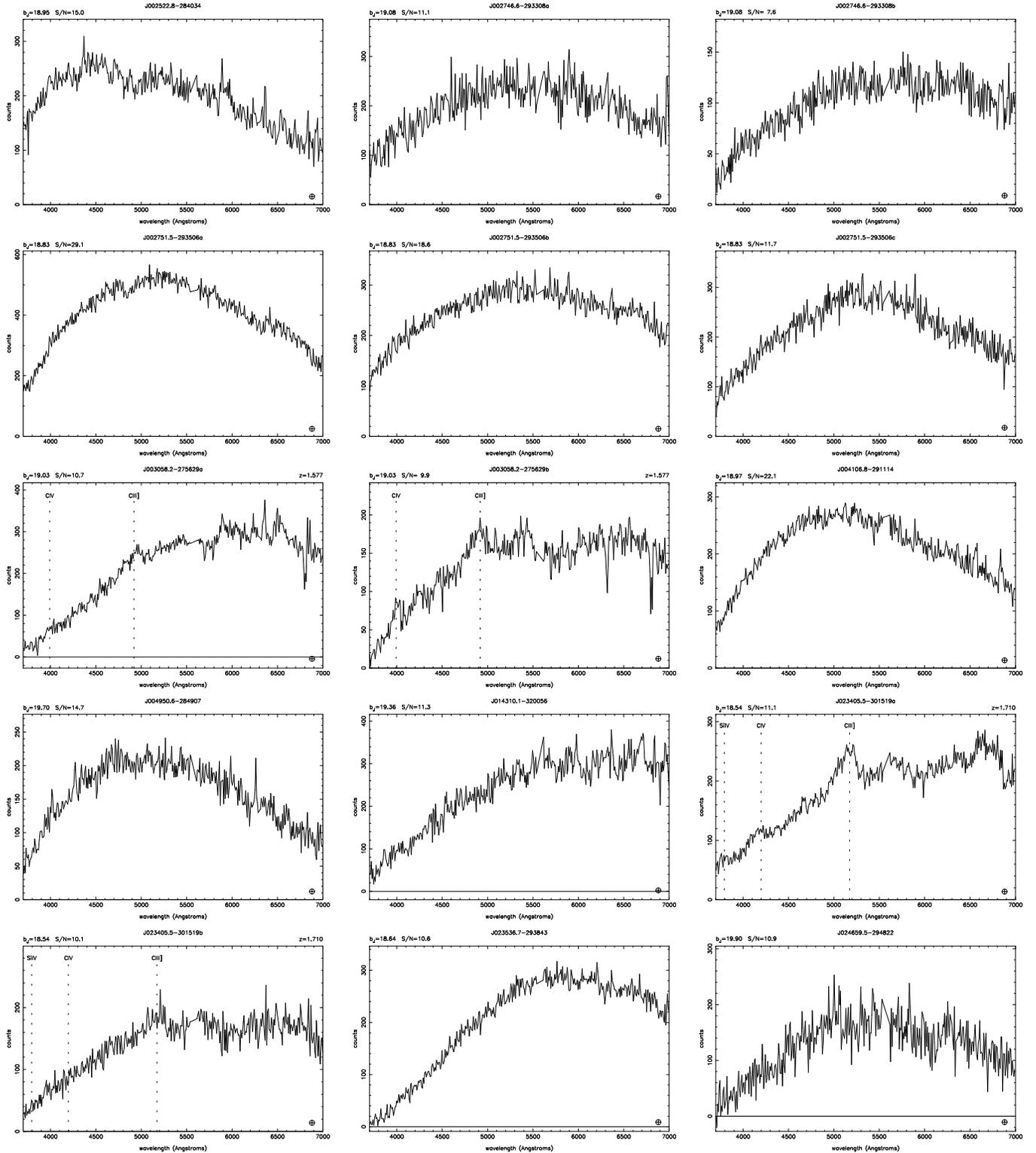}
\end{figure*}
\begin{figure*}
\psfig{file=bl_spec2.ps,width=18cm}
\end{figure*}
\begin{figure*}
\psfig{file=bl_spec3.ps,width=18cm}
\end{figure*}
\begin{figure*}
\psfig{file=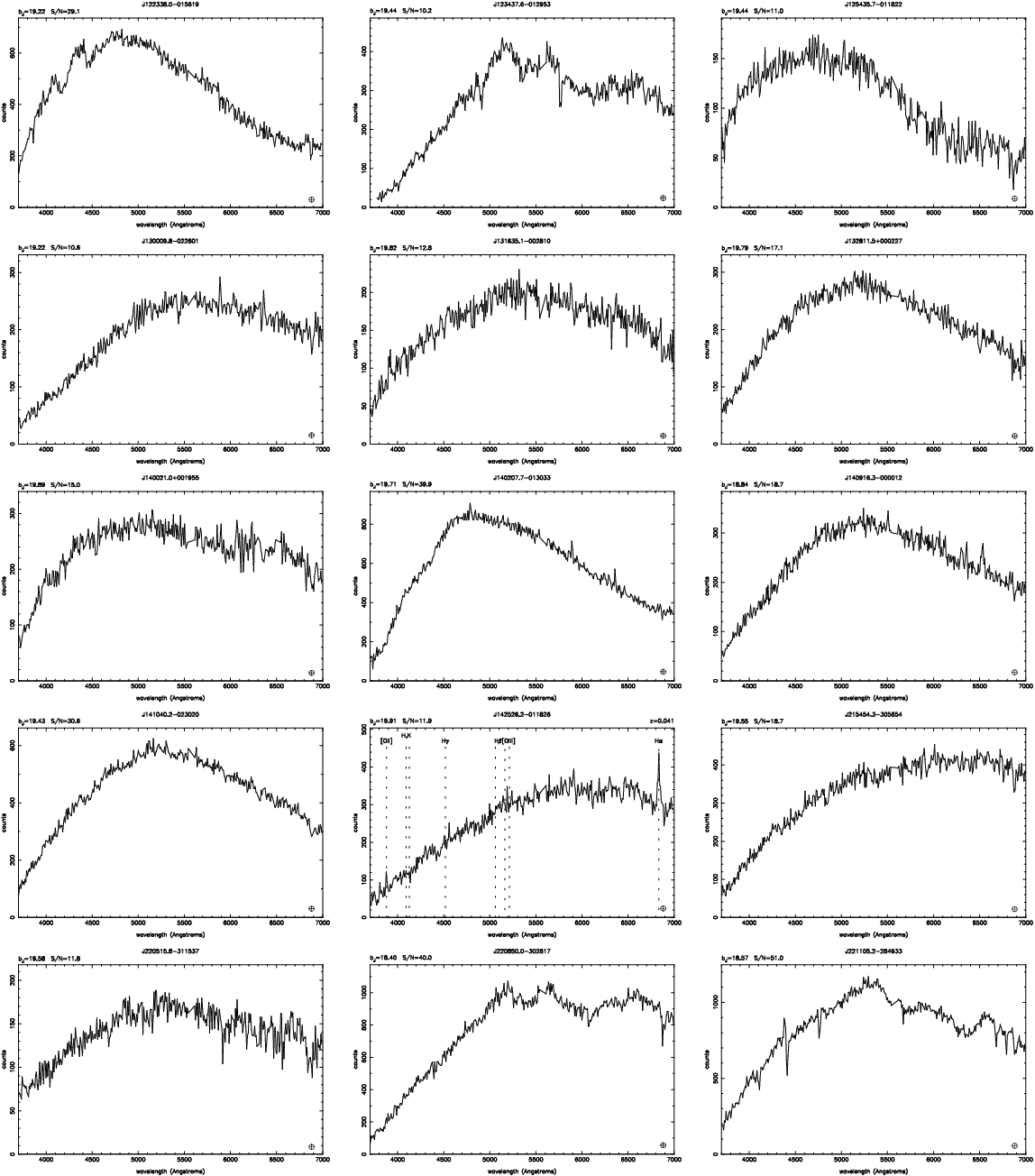,width=18cm}
\end{figure*}
\begin{figure*}
\psfig{file=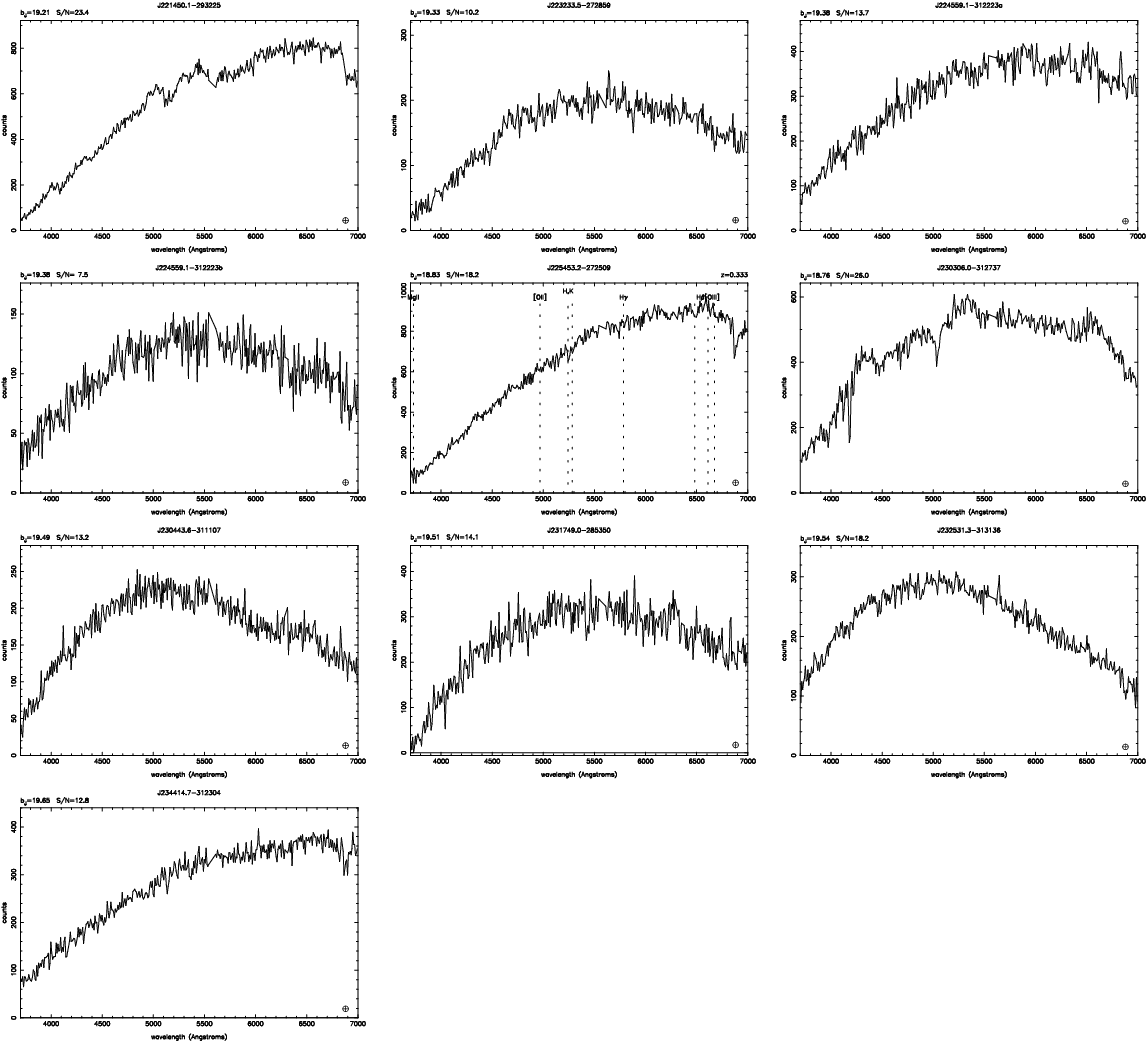,width=18cm}
\end{figure*}


\begin{thebibliography}{99}

\bibitem{1} Bade N., Beckmann V., Douglas N.G., Barthel P.D., Engles
D., Cordis L., Nass P. Voges W., 1998, A\&A, 334, 459\
\bibitem{2} Bailey J. et al., 2002, MNRAS, submitted\
\bibitem{3} Becker R.H., Gregg M.D., Hook I.M., McMahon R.G., White
R.L., Helfand D.J., 1997, ApJ, 479, L93\
\bibitem{4} Boyle B.J., 1989,  MNRAS, 240, 533\
\bibitem{5} Boyle  B.J., Shanks T., Croom S.M., Smith R.J., Miller L.,
Loaring N., Heymans C., 2000, MNRAS, 317, 1014\
\bibitem{6}Brotherton M.S.,  van Breugel W., Smith R.J., Boyle B.J., Shanks T., Croom S.M., Miller L., Becker R.H., 1998, ApJL, 505, 7 \
\bibitem{7}Browne I.W.A, ,  March\~{a} M.J.M., 1993, MNRAS, 261, 795\
\bibitem{8}Condon J.J., Cotton W.D., Greisen E.W., Yin Q.F., Perley
R.A., Broderick J.J., 1998, AJ, 115, 1693\
\bibitem{9}Corbett E.A., Robinson A., Axon D.J., Hough J.H., 2000,
MNRAS, 311, 485\
\bibitem{10} Croom S.M., Smith R.J., Boyle B.J., Shanks T., Loaring
N.S., Miller L., Lewis I.J., 2001, MNRAS, 322, L 29\
\bibitem{11}Fan X., et al., 1999, ApJ, 526, L57\
\bibitem{12} Fanaroff B.L., Riley J.M., 1974 MNRAS, 167, 31\
\bibitem{13}Fosatti G., 2001, in Padovani P. and Urry M., eds., ASP Conf. Ser. Vol 227 Blazar Demographics and Physics. Astron. Soc. Pac., San Francisco, p.218\
\bibitem{14} Gehrels N., 1986, ApJ, 303, 336 \
\bibitem{15} Giommi P., Pellizzoni A., Perri M., Padovani P., 2001, in
Padovani P. and Urry M., eds., ASP Conf. Ser. Vol 227 Blazar Demographics and Physics. Astron. Soc. Pac., San Francisco, p.227\
\bibitem{16}Laurent-Muehleisen S.A., Kollgaard R.I., Fiegelson E.D.,
Brinkmann W., Sibert J., 1999, ApJ, 525, 127 \
\bibitem{17}Lewis I. J. et al., 2001, MNRAS, submitted (astro-ph 0202175)
\bibitem{18}Maccacaro, T., della Ceca, R., Gioia, I.M., Morris, S.L.,
Stocke, J.T., Wolter, A., 1991, ApJ, 37,. 117\
\bibitem{19}McCook G.P., Sion E.M., 1999, ApJS, 121, 1\
\bibitem{20}March\~{a} M.J.M., ,  Browne I.W.A., 1995, MNRAS, 275, 951\
\bibitem{21}Miller, L., Peacock, J.A., Mead, A.R.G., 1990, MNRAS, 244, 207\
\bibitem{22}Morris et al., 1991, ApJ, 380, 49\ 
\bibitem{23}Outram P.J., Smith R.J., Shanks T., Boyle B.J., Croom S.M., Loaring
N.S., Miller L., 2001, MNRAS, 328, 805\
\bibitem{24}Padovani P.,  Giommi P., 1995 ApJ, 444, 567\
\bibitem{25}Rector T.A.,Stocke J.T., Perlman E.S.,Morris S.L., Gioia I.M., 2000, AJ, 120, 1626\
\bibitem{26}Rector T.A., Stocke J.T., Perlman E.S., 1999, ApJ, 516, 145\ 
\bibitem{27}Rector T.A. ,  Stocke J.T., 2001, AJ, 122, 565\
\bibitem{29}Sault R.J., Teuben P.J. Wright, M C H., 1995, PASP Conf. S\
\bibitem{30}Sault R.J., Killeen N.E.B., 1999, Miriad Users's Manual, http://www.atnf.csiro.au/computing/software/miriad\
\bibitem{31}Sion E.M., Fritz M.L., McMullin J.P., Lallo, M.D. 1988,
AJ, 96, 251\
\bibitem{32}Smith R.J., Croom S.M., Boyle B.J., Shanks T.,  Miller L., Loaring N.S. 2002, MNRAS, submitted (Paper III)\
\bibitem{33}Stickel M., Padovani P., Urry C.M., Fried J.W., Kuhr H.,
1991, ApJ, 374, 431\
\bibitem{34}Stocke, J.T., Morris, S L., Gioia, I, Maccacaro, T.,
Schild, R. E., Wolter, A., 1990, ApJ, 348, 141 \
\bibitem{35}Urry C.M.,  Padovani P., 1995, PASP 107, 803\
\bibitem{36}Voges W. et al., 1999, A\&A, 349, 389\
\bibitem{37}Wolter A., Caccianiga A., Della Ceca R., Maccacaro,T.,
1994, ApJ, 433, 29\
\bibitem{38}Wesemael, F., Greenstein, J.L., Liebert, J., Lamontagne,
R., Fontaine, G., Bergeron, P., Glaspey, J.W., 1993, PASP, 105, 761\
\bibitem{39}White R.L.,Becker R.H., Helfand D.J., Gregg M.D., 1997,
ApJ, 475, 479\
\end{thebibliography}
\end{document}